%% file: main.tex
\documentclass[11pt,a4paper]{article} 

\usepackage[numbers]{natbib}  
\usepackage{graphicx}

\usepackage[utf8]{inputenc} 
\usepackage[T1]{fontenc}
\usepackage{amsmath,amssymb,amsfonts} 
\usepackage{graphicx} 
\usepackage{hyperref} 
\usepackage{geometry} 
\usepackage{caption}
\usepackage{subcaption} 

\usepackage{comment}

\usepackage{float}

\usepackage{xcolor}

\usepackage{siunitx}  

\usepackage{multirow}

\usepackage{float}%
\usepackage{caption}
\usepackage{amssymb}
\usepackage{amsmath}
\usepackage{esint}

\title{Wave-like Spiral Interactions and the Emergence of Quasi-Biennial Oscillations in Strato-Rotational Flows}

\author{
  G. Meletti\thanks{Departament de Matemàtiques, Universitat Politècnica de Catalunya, Av. Diagonal 647, 08028 Barcelona, Spain}
  \and
  S. Abide\thanks{D\'epartement de Math\'ematiques, Universit\'e C\^ote d'Azur, Parc Valrose 06108 Nice CEDEX 2, France}
  \and
  S. Viazzo\thanks{Laboratoire de M\'ecanique, Mod\'elisation et Proc\'ed\'es Propre, Aix-Marseille University, CNRS, 38 rue Frédéric Joliot Curie, 13451 Marseille, France}
  \and
  J. Curbelo\thanks{Centre de Recerca Matemàtica, Campus de Bellaterra, Edifici C, 08193 Bellaterra, Barcelona, Spain}
  \footnotemark[1] 
  \and
  U. Harlander\thanks{Department of Aerodynamics and Fluid Mechanics, Brandenburg University of Technology (BTU) Cottbus-Senftenberg, Siemens-Halske-Ring 14, 03046 Cottbus, Germany}%
  \thanks{Corresponding author: \texttt{uwe.harlander@b-tu.de}}
}

\date{\today}

\begin{document}
\maketitle

\begin{abstract}

This study investigates the dynamics of Strato-Rotational Instability (SRI) in a stratified, rotating fluid, focusing on the interaction between axial modes and spiral components. Through numerical analysis, we find that SRI induces oscillatory behaviors that change the mean flow, leading to the selective activation of distinct axial wavenumbers associated with upward and downward propagating spiral modes. These results suggest wave-mean flow interactions. The use of Radon Transforms (RT) allowed us to separate these spiral components, showing that each upward and downward component was individually modulated, but out of phase with each other. Inspired by the RT findings, a simplified toy model was developed to interpret the spiral pattern changes linked to amplitude modulations. The model considers two wave-like spirals propagating in opposite axial directions, linearly interacting. By incorporating out-of-phase individual spiral modulations, the model reproduces the observed spiral pattern transitions, offering a straightforward interpretation of the underlying physical processes. To explore the mechanism of individual spiral modulations, we consider a Quasi-Biennial Oscillation (QBO)-like framework derived from the Navier–Stokes equations in a rotating frame. These findings contribute to a better understanding of low-frequency SRI dynamics and may offer insights into similar phenomena in geophysical and astrophysical contexts.

\end{abstract}

\section*{Keywords}
Rotating flow, Stratified flow, Waves in rotating fluids, Pattern formation, Particle Image Velocimetry, High Performance Computing, Direct Numerical Simulations.


\newpage

\input{content/introduction}

\input{content/NumMethods}

\input{content/OLDSpiralMode}
 \input{content/ToyModel}

 \input{content/AxialMeanFlow}
 \input{content/conclusion}

\section*{Acknowledgements}

The authors also thank A. Krebs, T. Seelig, A. Randriamampianina, and I. Raspo for the discussions and support. We also thank the researchers from the Nonlinear Fluid Dynamics of the Universitat Polit\`{e}cnica de Catalunya for the constructive discussions. 
 Uwe Harlander acknowledges support from the DAAD project ``Combined studies of baroclinic waves with methods of data assimilation'' (57560889). Gabriel Meletti acknowledge the financial support from the DFG core facility center ’Physics of rotating fluids’, DFG HA 2932/10-1. Gabriel Meletti and Jezabel Curbelo acknowledge support from  the Agencia Estatal de Investigación through the grant  CNS2023-144360 funded by the “European Union NextGenerationEU/PRTR”. She also thanks the Severo Ochoa and María de Maeztu Program for Centers and Units of Excellence in R\&D (CEX2020-001084-M) and the Fundación Ramón Areces. Stéphane Abide has been supported by the French government, through the UCAJEDI Investments in the Future project managed by the National Research Agency (ANR) with the reference number ANR-15-IDEX-01.


\bibliographystyle{plainnat}  
\bibliography{SRI}

\input{content/Appendices}

\end{document}

%% file: content/introduction.tex
\section{Introduction}\label{sec:introduction}

The interaction between rotation and stable density stratification is frequently found in geophysical and astrophysical flows, often leading to instabilities that play a crucial role in the development of large-scale circulation patterns observed in nature. In geophysical systems such as the Earth's atmosphere and oceans, stratified rotating flows significantly influence weather and climate patterns. A well-known phenomenon in these systems is the Quasi-Biennial Oscillation (QBO), observed in Earth's equatorial stratosphere. The QBO is characterized by a periodic reversal of zonal winds, driven by the emission and interaction of inertia-gravity waves (IGWs) with the mean flow \citep{lindzen1968theory, plumb1977interaction, Plumb_McEwan}. Understanding the mechanisms underlying the QBO provides valuable insights into the broader dynamics of rotating stratified flows and highlights the importance of wave-mean flow interactions \citep{seelig2015can}.

In astrophysical contexts, the interaction between rotation and stable density stratification can be the key to understanding and modeling the formation and evolution of planetary systems that form from accretion disks, as they facilitate the outward transport of angular momentum, allowing matter to aggregate under gravitational forces \citep{fromang2017angular, lyra2019initial}. Accretion disks are gas and dust structures that have differential Keplerian (or near-Keplerian) velocity profiles \citep{visser2010sub} that can be approximated as a Taylor-Couette (TC) system, which consists of a fluid confined between two independently rotating concentric cylinders. TC systems then serve as a canonical model for investigating the fundamental behaviors of astrophysical rotating flows \citep{Dubrulle2004}. When stable density stratification in the axial direction is introduced to a classic TC system, a purely hydrodynamic instability known as strato-rotational instability (SRI) can develop \citep{Withjack, Boubnovt1995, LeBars2007}. This instability manifests as non-axisymmetric spiral modes and has gained attention as a potential mechanism for enhancing angular momentum transport in accretion disks. The SRI can destabilize flow regimes that would otherwise be stable under non-stratified conditions, thereby providing a pathway for angular momentum transfer in accretion disks and similar environments \citep{Yavneh2001, Dubrulle2004, Shalybkov2005}.

While the SRI arises in the presence of rotation and density stratification, the underlying mechanisms that lead to its development are not yet fully understood. \citet{molemaker2001instability}, e.g., suggests that 
the SRI 
results from the interaction between Coriolis and pressure forces, leading to the superposition of two Kelvin waves that resonate in the boundaries of the system. In contrast, \citet{park2012waves,wang2018strato} argue that the SRI could emerge from the spontaneous radiation of internal waves, which reflect and resonate within the cavity, interacting with critical layers to generate the instability, independently of the presence of a rigid outer boundary \citep{Dizes2010}. 
%
%

\begin{figure}
\centering
	\begin{minipage}[t]{0.4\linewidth}
		\centering
		\includegraphics[width=0.92\linewidth]{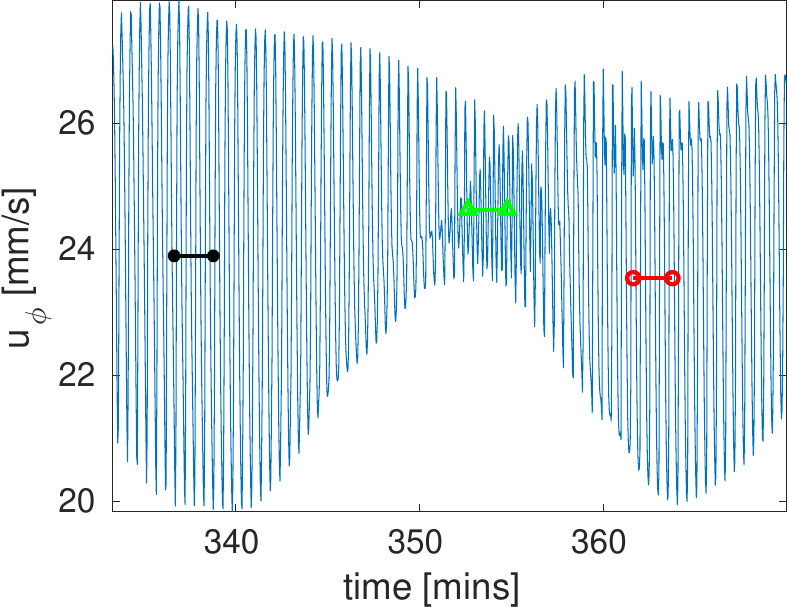}
		\caption*{{(a)} $u_\phi$ time series} 
		\vspace{2ex}
	\end{minipage}
	\begin{minipage}[t]{0.4\linewidth}
		\centering
		\includegraphics[width=1\linewidth]{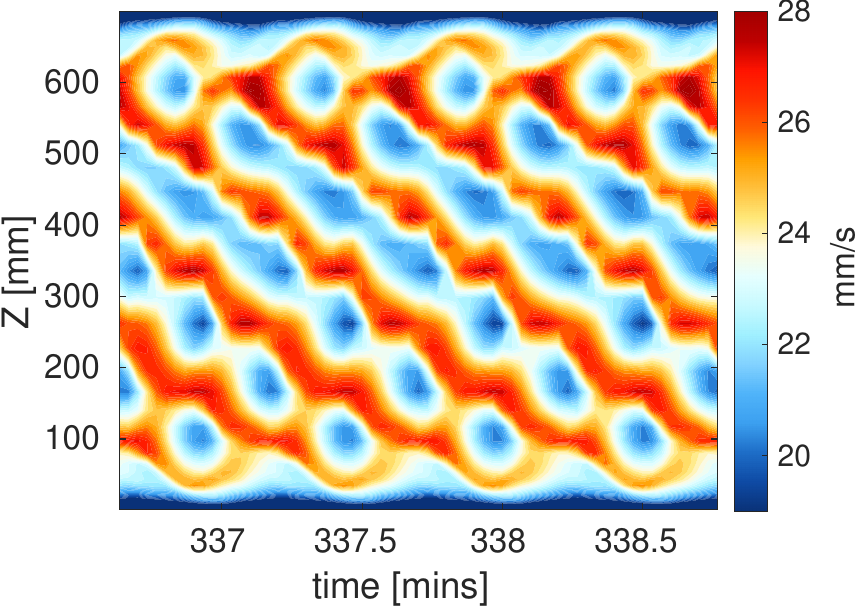}
		\caption*{{(b)} (black) Interval 01} 
		\vspace{2ex}
	\end{minipage} 
	\begin{minipage}[t]{0.4\linewidth}
		\centering
		\includegraphics[width=1\linewidth]{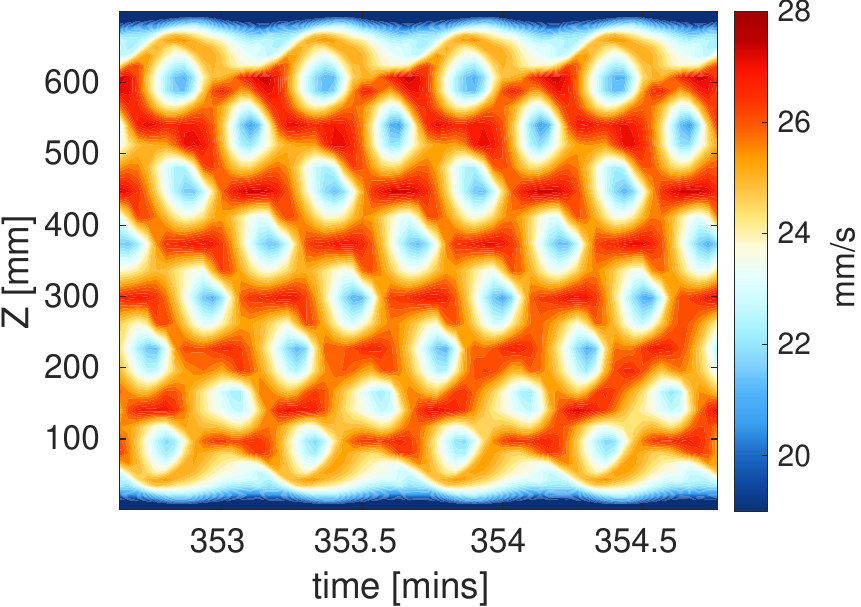}
		\caption*{{(c)} (green) Interval 02} 
		\vspace{2ex}
	\end{minipage}
	\begin{minipage}[t]{0.4\linewidth}
		\centering
		\includegraphics[width=1\linewidth]{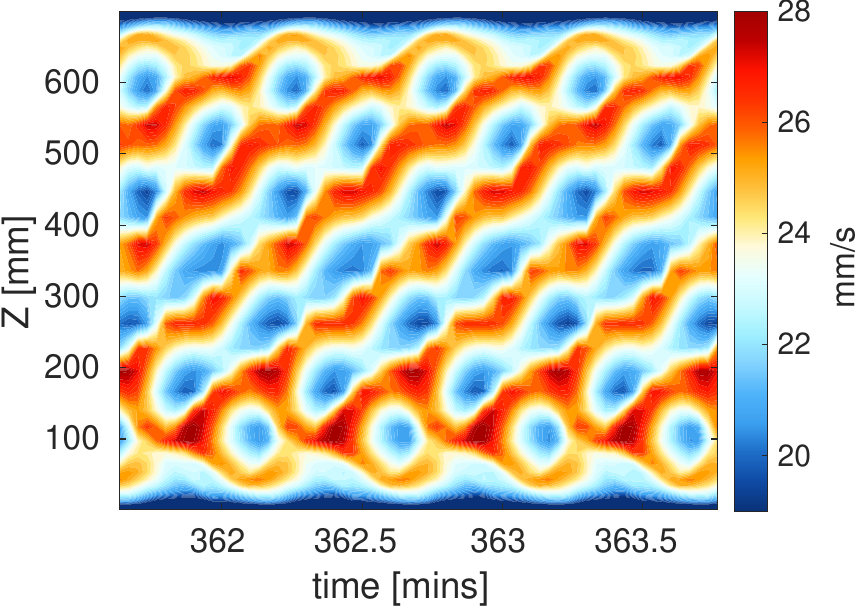}
		\caption*{{(d)} (red) Interval 03}
		\vspace{2ex}
	\end{minipage}
	\caption{$u_\phi$ structures during amplitude modulation at radial position $r=r_{in}+d/2$, inspired by \citet{Meletti2020GAFD}; (a) Time series with horizontal colored lines indicating intervals selected before (black), during (green), and after (red) a local minimum amplitude value; (b)~Interval 01, from t$\approx$~336 to 339 minutes -- SRI spiral with downward inclination; (c)~Interval 02, from t$\approx$ 352 to 355 minutes -- transition from a SRI spiral with downward to upward inclination; (d)~Interval 03, from t$\approx$ 361 to 364 minutes -- SRI spiral with upward inclination.}
	\label{fig:Pattern_modulation}
\end{figure}

Recent studies have shown that the SRI can lead to intriguing spiral pattern changes associated with low-frequency velocity modulations observed both numerically and experimentally \citep{Meletti2020GAFD, lopez2022stratified}. In \citet{Meletti2020GAFD}, distinct flow patterns along the axial direction were linked to these modulations. Figure~\ref{fig:Pattern_modulation}, inspired by~\citet{Meletti2020GAFD}, shows the patterns associated with amplitude modulations in three distinct time intervals. Each interval displays a unique flow pattern in the space-time diagrams (axial-time frame), which will be further explored in this study. The downward-inclined spiral, presented in figure~\ref{fig:Pattern_modulation}(b), travels from the top lid to the bottom in the axial direction, while the upward-inclined spiral, shown in figure~\ref{fig:Pattern_modulation}(d), moves in the upward direction. During the transition phase, illustrated in figure~\ref{fig:Pattern_modulation}(c) and characterized by a chessboard-like structure and smaller SRI amplitudes (figure~\ref{fig:Pattern_modulation}(a)), on which the superposition of the two spirals leads to a standing
wave pattern. 

The observed SRI spirals share similarities with those found in experiments by \citet{flor2018onset}, where two spirals move in opposite directions, forming a standing pattern, and with simulations presented by \citet{lopez2020impact}, with low-frequency modulations due to differences in their axial drift speeds. However, it is important to note that \citet{flor2018onset} and \citet{lopez2020impact} utilized a short annulus with a wide gap, where top and bottom boundary effects and centrifugal buoyancy play significant roles. Additionally, unlike the studies by \citet{Meletti2020GAFD, meletti2023parameter} and the present paper, \citet{lopez2020impact} employed a smaller Froude number ($Fr < 1$) and a larger Reynolds number ($Re > 6000$). In longer (taller) cavities, spiral pattern changes in the axial-radial plane were also observed experimentally by \citet{Riedinger2010}.


This present study aims to investigate the mechanisms driving spiral pattern changes in stratified rotating flows. We propose that these spiral pattern changes can be modeled as two individually modulated wave-like spirals, superposed with a phase shift between them. After observing the presence of weak non-linear features in the modulation of each spiral, we link the mechanism of modulation and pattern changes to quasi-biennial oscillation (QBO)-like behavior. 

The paper is structured as follows: Section~\ref{sec:NumMethods} describes the numerical methods used in our study, detailing the direct numerical simulation (DNS) approach, the governing equations, and the boundary conditions. In Section~\ref{sec:radonSeparation}, we examine variations in the base flow due to the strato-rotational instability (SRI), focusing on the mode activation during spiral pattern changes using 2D-Fast Fourier Transform (2D-FFT) analysis. We also investigate the behavior of spiral components by applying the Radon Transform (RT) to separate upward and downward traveling spirals. Section~\ref{sec:ToyModel} introduces a simplified toy model illustrating wave-like spiral propagation to explain the observed pattern transitions. Section~\ref{sec:AxialMeanFlow} proposes a QBO-like model to explain the origin of the amplitude modulations. Finally, Section~\ref{sec:conclusion} concludes the paper by summarizing the key findings.

%% file: content/NumMethods.tex
\section{Numerical Methods}\label{sec:NumMethods}

In this paper, we examine SRI amplitude modulations linked to spiral pattern changes in a Taylor-Couette system subjected to heating from above and cooling from below, which establishes a stable density stratification along the axial ($z$) direction (i.e., with less dense fluid on top, and more dense fluid as we move towards the bottom of the cylindrical cavity along its axial(z)-axis).

 We will use the direct numerical simulation (DNS) solver developed by \citet{ABIDE2018}, considering the same physical model, numerical methods, and experimental parameters as we used in \cite{seelig2018experimental,Meletti2020GAFD,meletti2023parameter}, so that the results can be consistent with previous findings. The code employs a fourth-order accurate spatial discretization and high-performance computing (HPC) \citep{ABIDE2017,ABIDE2018}. 

 The physical model adopts a Taylor-Couette flow configuration filled with an incompressible fluid. The fluid properties considered are those of the M5 silicon oil used in the experiments presented in \citet{seelig2018experimental, Meletti2020GAFD}, with kinematic viscosity $\nu = 5\times10^{-6}{\mathrm m}^2/{\mathrm s}$; specific weight $\rho = 923{\mathrm {kg}}/{\mathrm m}^3$;  coefficient of thermal expansion $\alpha=1.08\times10^{-3}/{\mathrm K}$; thermal conductivity $k=0.133 {\mathrm W}/{\mathrm {K m}}$; specific heat $c_p = 1630 {\mathrm J}/{\mathrm {kg K}}$; and Prandtl number $Pr =  57$.  
 
 A vertical temperature gradient of $\Delta T/\Delta z = \SI{5.71}{\kelvin\per\metre}$ ensures the stable axial density stratification. The geometrical parameters of the TC cylindrical cavity are presented in table~\ref{table:Exp_parameters}.

\begin{table}
\begin{center}
\caption{Setup geometrical parameters \label{table:Exp_parameters}}
\begin{tabular}{lcc}
\hline
    inner cylinder radius & $r_{in}$ & $75{\mathrm {mm}}$ \\ 
    outer cylinder radius & $r_{out}$ & $145{\mathrm {mm}}$ \\
    gap size & $d$ &  $70{\mathrm {mm}}$ \\ 
    cylinders height & $H$ & $700{\mathrm {mm}}$ \\
    aspect ratio & $\varGamma$ & $10$ \\
    radii ratio & $\eta$ & $\approx  0.52$  \\ 
\hline
\end{tabular}
\end{center}
\end{table}
 
 The numerical code applied here solves the Navier-Stokes equations using the Boussinesq approximation to incorporate buoyancy forces. The governing equations for the code then read:

\begin{equation}\label{eq:Numeric_Eqn}
\left\{
\begin{array}{ll}
\nabla \cdot {\textbf{u}} = 0 & \text{ in }  D, \\
\displaystyle \partial_t \textbf{u} + \frac{1}{2}\left[(\textbf{ u} \cdot \nabla ) \textbf{ u} + \nabla \cdot (\textbf{ u}\textbf{ u}) \right] = - \nabla p + \nu \Delta \textbf{ u} +\mathbf{F} & \text{ in }  D,\\
\displaystyle \partial_t T + \frac{1}{2}\left[(\textbf{ u} \cdot \nabla ) T+ \nabla  (\textbf{ u} T) \right] = \kappa \nabla^2 T & \text{ in }  D.\\
\end{array}
\right.
\end{equation}

Here, $D$ denotes the computational domain, $\kappa$ is the fluid's thermal conductivity, $p$ is the pressure, $T$ represents the temperature field, and $\textbf{ u} = (u_r, u_\phi, u_z)$ denotes the velocity vector field in the radial, azimuthal, and axial directions, respectively. The buoyancy force, denoted by $\mathbf{F}$, results from density variations and is defined as:

\begin{equation}
\mathbf{F} = \alpha \textbf{ g} \frac{\partial T}{\partial z}\Delta z .
\end{equation}

All analysis will consider intermediate Reynolds numbers of $Re = 400$, with rotation ratio between inner and outer cylinders $\mu = \Omega_{out}/\Omega_{in} = 0.35$. This value of $\mu$ is considered so that we impose a velocity profile that is almost Keplerian (slightly slower), to focus on accretion disk applications (see \citet{visser2010sub, lyra2019initial}). 

In our simulations, velocity boundary conditions are applied at the cylinder walls, with the top and bottom boundaries rotating at the same angular speed as the outer wall. To impose the temperature gradient, the temperature at the top and bottom lids are prescribed as adiabatic Dirichlet boundary conditions. Assuming negligible heat loss through the lateral walls compared to the thermal forcing at the lids, adiabatic boundary conditions are applied laterally, at the inner and outer cylinder's walls. Simulations begin with small, randomly distributed white noise perturbations to trigger instability development. Time discretization is conducted using the methods proposed by \citet{hugues1998improved}. Spatial discretization is enhanced using spectral Fourier methods in the azimuthal direction and a fourth-order compact finite difference scheme in the radial and axial directions, as described by \citet{ABIDE2005}. High-Performance Computing techniques, as detailed in \citet{ABIDE2017}, facilitate significant reductions in simulation time through parallelization. Following \citet{lopez2020impact} and \citet{lopez2022stratified}, centrifugal buoyancy effects did not present a strong influence in our simulations once the instability was already established, and they were therefore omitted from the numerical model. The simulations employ a grid resolution of $32 \times 64 \times 200$ in the $\phi \times r \times z$ directions.

%% file: content/OLDSpiralMode.tex
\section{Axial modes activation and spiral components separation}\label{sec:radonSeparation}

In this section, we will present the interaction of the SRI with the base flow, and how this is related to the spiral pattern changes. We will further discuss how this is connected to the activation of different axial wavenumbers. Moreover, we discuss how the oscillating behavior of the spiral axial propagation can be separated into two components, each of them related to weak non-linear features.

The initial detailed comparison of experimental and numerical data by \citet{Meletti2020GAFD} identified low-frequency amplitude modulations, as those shown here in figure~\ref{fig:Pattern_modulation}. Similar patterns were later obtained numerically by \citet{lopez2022stratified} for a short cylinder geometry. These amplitude modulations occur across all velocity components ($u_\phi$, $u_r$, and $u_z$) and the temperature $T$, affecting thus the temporal variations of the circulation and the stratification.

\begin{figure}
\begin{center}
	\begin{minipage}[t]{0.42\linewidth}
		\centering
		\includegraphics[width=1\linewidth]{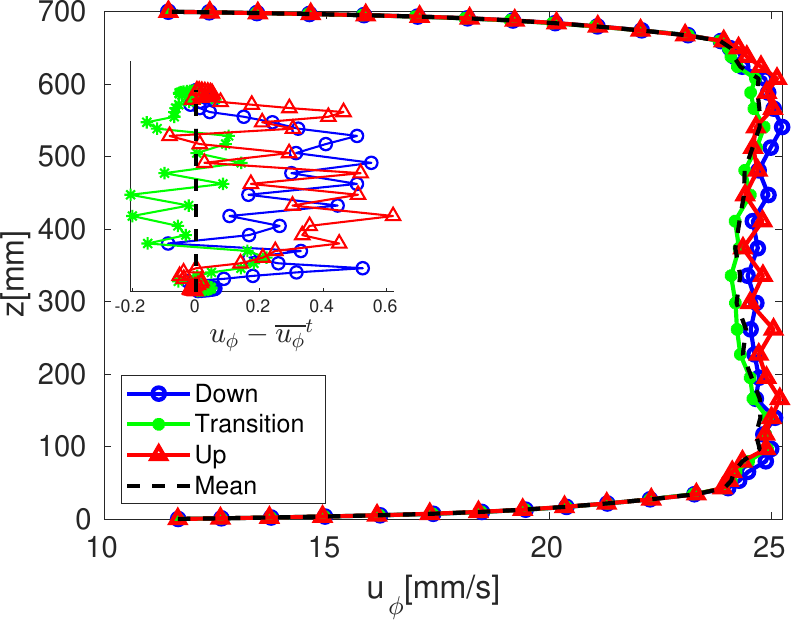}
		\caption*{{(a)~$u_\phi$}}
	\end{minipage}
    \begin{minipage}[t]{0.42\linewidth}
		\centering
		\includegraphics[width=1\linewidth]{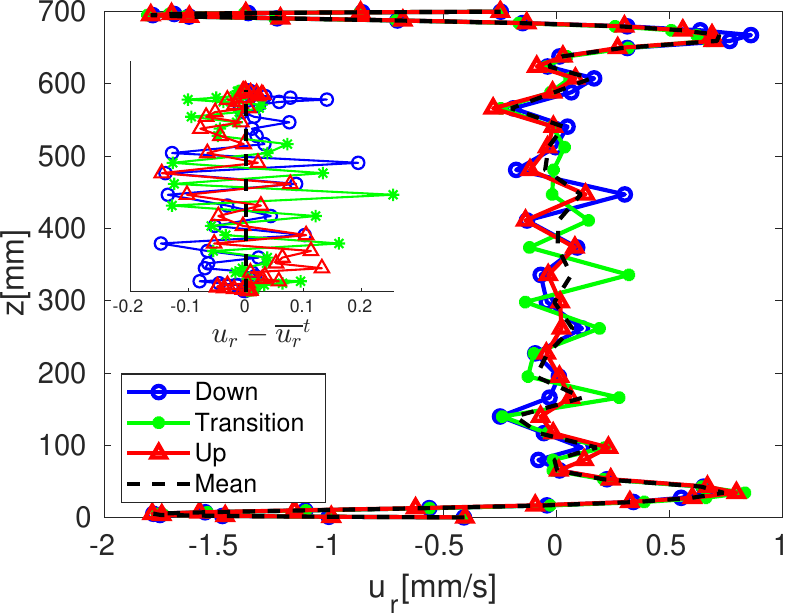}
		\caption*{{(b)~$u_r$}}
	\end{minipage}
	\begin{minipage}[t]{0.42\linewidth}
		\centering
		\includegraphics[width=1\linewidth]{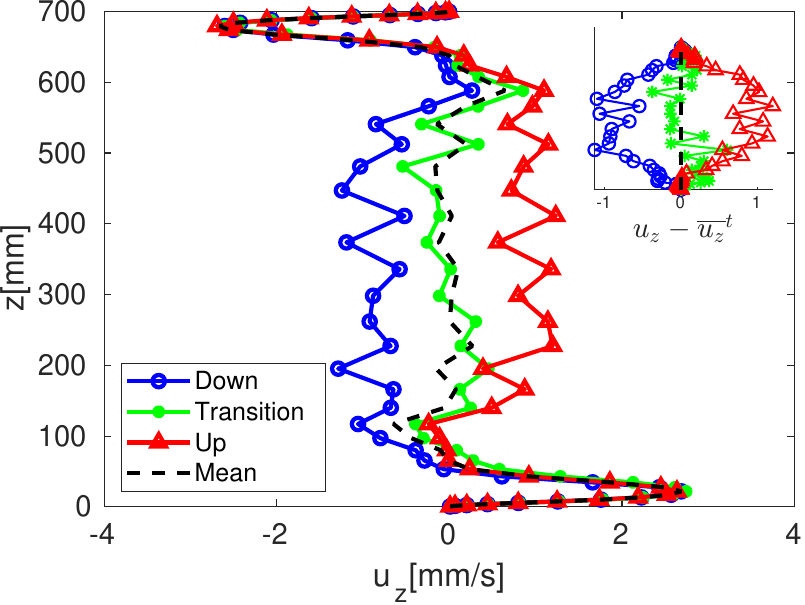}
		\caption*{{(c)~$u_z$}}
		\vspace{2ex}
	\end{minipage}	    
	\begin{minipage}[t]{0.42\linewidth}
		\centering
		\includegraphics[width=1\linewidth]{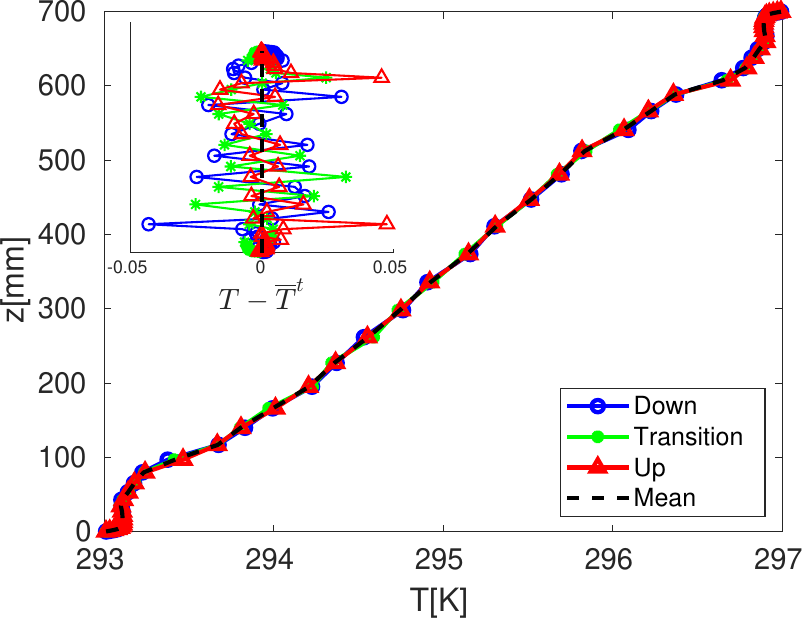}
		\caption*{{(d)~Temperature}}
		\vspace{2ex}
	\end{minipage}
	\caption{ Comparison of the time average axial velocity profiles with time averages taken during an upward traveling spiral period, during a downward traveling spiral period, and during the transition from an upward to a downward spiral period. The black dashed line shows the time average over four full periods of amplitude modulations, comprehending upward, downward, and standing spiral patterns. The results are from numerical simulation performed with $Re=400$, $\mu=0.35$ and $\Delta T/ \Delta z = \SI{5.71}{\kelvin\per\metre}$ at a fixed radial position $r = r_{in}+d/2$. The figures inserted inside each image show values of $f-\overline{f}^t$, i.e., the mean values during the upward, downward, and transition spiral traveling regime minus the average during the full simulation period.}
	\label{fig:MeanPerRegion}
	\end{center}
\end{figure}

 
 Figure~\ref{fig:MeanPerRegion} shows the time-averaged velocity profiles in the axial direction at a fixed radial position $r = r_{in} + d/2$. The plots inserted in each figure shows $u-\overline{u}^t$, i.e., the velocity minus the total mean value (which is represented in black dashed line). The time-averaged azimuthal velocity ($\overline{u_\phi}$), displayed in figure~\ref{fig:MeanPerRegion}(a), shows that during upward or downwards traveling regime (when the velocity amplitude of is larger),  the azimuthal flow is slightly accelerated with respect to the mean values or compared to the $u_\phi$ during the spiral transition between these two regimes. The radial velocity profiles ($\overline{u_r}$) presented in figure~\ref{fig:MeanPerRegion}(b), instead, do not change significantly during the periods when the spirals travel either upwards or downwards, showing no great differences from the overall time average. However, figure~\ref{fig:MeanPerRegion}(c) illustrates how the time-averaged axial velocity ($\overline{u_z}$) is affected by the instability. The mean axial flow becomes positive during upward spiral propagation and negative during downward spiral propagation. During the transition between the two propagation patterns, the average axial velocity approaches zero. These variations in the time-averaged axial velocity are important because they show that the spiral propagation affects the mean flow.  This connection highlights the interaction between the base flow and the emerging instabilities. Figure~\ref{fig:MeanPerRegion}(d) shows the temperature profiles (and their fluctuations). In these cases, we see some difference in temperature fluctuations, of about 0.5$^oC$, especially near the cavity lids, which are considered to be small.
Note that near the top and bottom boundaries in figure~\ref{fig:MeanPerRegion}, there is a strong radial inward flow caused by Ekman circulation, with a strong upward flow near the bottom wall and a downward flow near the top wall (shown in the radial velocity presented in figure~\ref{fig:MeanPerRegion}(c)). These Ekman effects are qualitatively similar to those reported by \citet{lopez2020impact}. Despite these Ekman effects, simulations with periodic boundary conditions at the top and bottom lids (not presented here) also exhibited amplitude modulations associated with pattern changes (see~\citet{meletti2023parameter}), showing that the presence of lids and Ekman effects is not the reason for the modulations and pattern transitions to occur.

\begin{figure}
\begin{center}
	\begin{minipage}[t]{0.32\linewidth}
		\centering
		\includegraphics[width=1\linewidth]{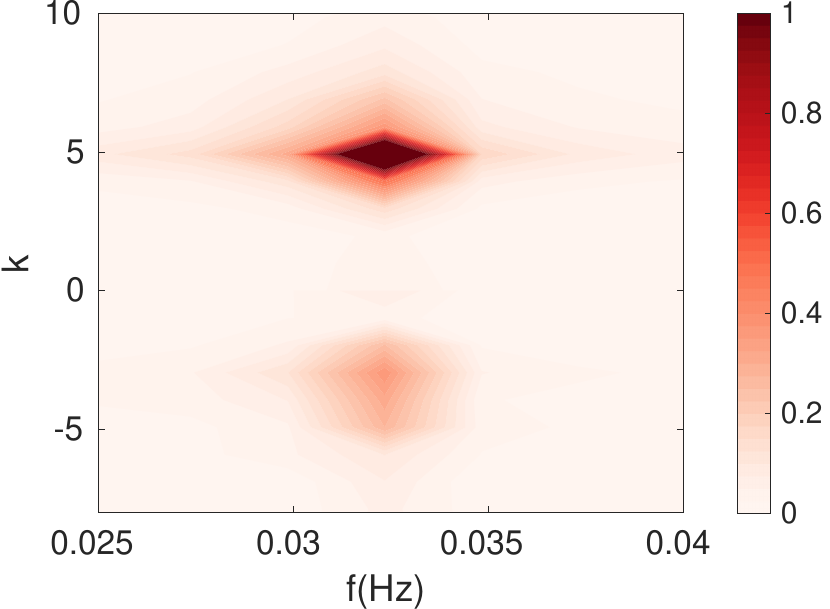}
		\caption*{{(a)}}
	\end{minipage}
    \begin{minipage}[t]{0.32\linewidth}
		\centering        
		\includegraphics[width=1\linewidth]{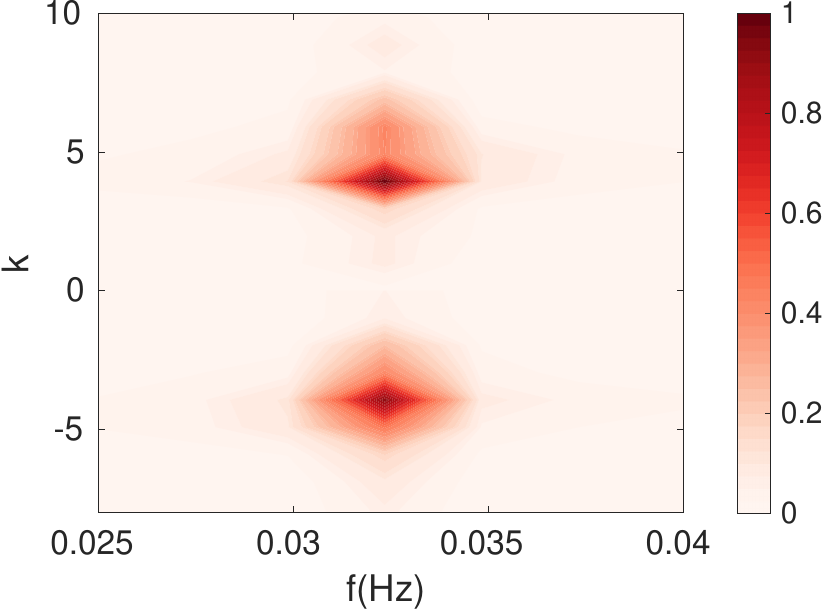}		
		\caption*{{(b)}}
	\end{minipage}	
	\begin{minipage}[t]{0.32\linewidth}
		\centering
        \includegraphics[width=1\linewidth]{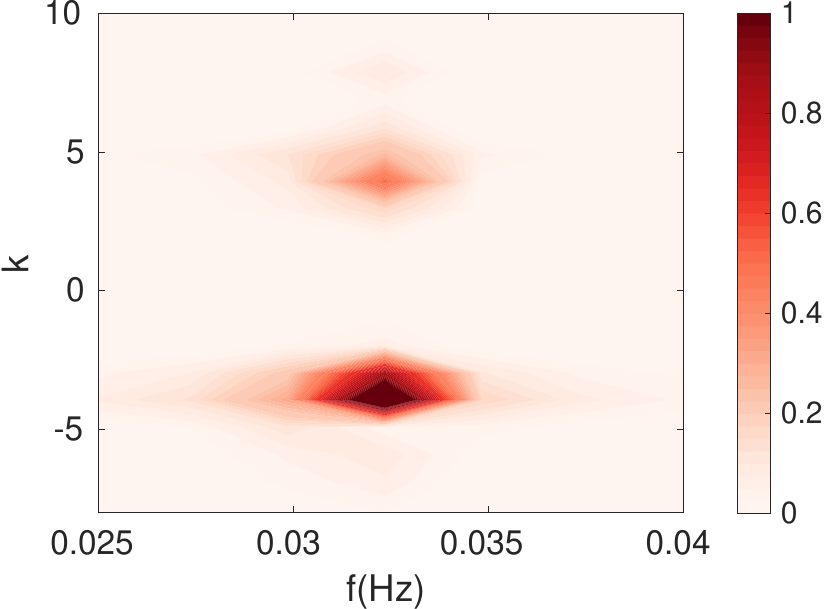}
		\caption*{{(c)}}
		\vspace{2ex}
	\end{minipage}
	\end{center}
	\caption{2-dimensional power spectra of $u_\phi$ in the axial direction during {(a)}~upward spiral propagation; {(b)}~transition from a downward an upward propagating spiral; {(c)}~downward spiral propagation. The $y$-axis represents axial wavenumber k (axial modes), while the $x$-axis is the frequency in $Hz$.  The spectra amplitudes are normalized by the maximum amplitude value of the upward (and downward) propagating spirals $P_{0,max}$. The spectra are obtained in a frame of reference fixed in the laboratory.  During the transition, the maximum amplitude of the spectra was half of the maximum amplitude found while the spiral was traveling upward or downward ($P/P_{0,max} = 0.5$).  }
	\label{fig:2D-FFT}
\end{figure}

To investigate if there are changes in the SRI frequency and in the axial wavenumbers during these different spiral propagating regimes, we computed the 2D-Fast Fourier Transform (2D-FFT) obtained from $u_\phi$ space-time diagrams in the axial direction (presented in figure~\ref{fig:Pattern_modulation}). Figure~\ref{fig:2D-FFT} shows the results obtained at moments when the spiral is traveling upwards (figure~\ref{fig:2D-FFT}(a), on the left-hand side), downwards (figure~\ref{fig:2D-FFT}(c), on the right-hand side), and during the transition from the upward to the downward propagation (figure~\ref{fig:2D-FFT}(b), middle image). The diagrams in figure~\ref{fig:2D-FFT} present the frequencies $f$ on the $x$-axis, and the axial wavenumber $k$ in the $y$-axis. 
We note that the peaks have the same SRI frequencies of $f=0.032$Hz  independently of the spiral direction of propagation, but they change their wavenumbers from positive to negative values, with different modes being activated and suppressed. 
During the upward traveling spiral, the stronger wavenumber activated in left-hand side figure~\ref{fig:2D-FFT}(a) is $k_{up}=4$, with maximum amplitude $P_{0,max}$, and the downward mode, which has wavenumber $k_{down}=-4$, is weakly activated. The wavenumber activated with $P_{0,max}$ in the right-hand side figure~\ref{fig:2D-FFT}(c), related to the downward propagating spiral, is $k_{down}=-4$, while $k_{up}=4$ has smaller amplitude. On the middle figure figure~\ref{fig:2D-FFT}(b), both modes $k_{up}=4$ and $k_{down}=-4$ are activated, each with approximately half of the maximum amplitude of the spiral traveling upward or downward ($P_{0,max}/2$). 

To better understand the spiral oscillatory behavior and its associated mode activation during the different spiral patterns observed, we separated the up$-$ and downward propagating spirals by applying the Radon Transform (RT) to the full signal, capturing phases during dominating up, downward, and mixed spiral propagation. The Radon transform is a Fourier-like technique to select wave components with different directions of propagation. The RT is particularly suited for finding individual waves that compose noisy or irregular fields \citep{almar2014use}. These techniques are interesting for evaluating the results directly using the data obtained, without knowing the wave's dispersion relation and without the necessity of an analytical model of the SRI. A brief description of the Radon transforms can be found in the appendix~\ref{appendix:Radon}.

Figure~\ref{fig:Radom_Re400} then shows the separation of the upward and downward components of $u_\phi$ on a space-time diagram while the spiral is traveling upward (left-hand side) and downward (right-hand side) using the RT. The results are from the simulation with $Re=400$, $\mu=0.35$, and $\Delta T/\Delta z \approx \SI{5.71}{\kelvin\per\metre}$ presented in figure~\ref{fig:Pattern_modulation}. After the two wave fields have been separated, we again computed the 2D-FFT spectra from the corresponding space-time diagrams (not shown here).
We observed that the 2D-FFT of the downward spiral component is exactly the same as the bottom wavenumber in the full 2D-FFT spectrum presented in figure~\ref{fig:2D-FFT}, but the positive frequency is no longer observed (as expected). The upward wavenumber in figure~\ref{fig:2D-FFT} was also captured in the spectrum of the upward component of the upward-traveling spiral (Fig.~\ref{fig:Radom_Re400}(e)), but the downward traveling spiral component is removed. Thus, on a simplified model, the wavenumbers we observe on figures~\ref{fig:2D-FFT} can be associated with a superposition of one upward and one downward spiral of axial wavenumber $k=4$ and $k=-4$ propagating in time with the SRI frequency (see \citet{Meletti2020GAFD} for more details on the SRI frequency measurements and values). In other words, while the spiral is propagating downwards, we see in figure~\ref{fig:Radom_Re400} that the spiral traveling upwards is suppressed, reaching smaller amplitudes and a more vertical inclination (figure~\ref{fig:Radom_Re400}(f)). The same occurs with the downward component when the spiral is propagating upwards (figure~\ref{fig:Radom_Re400}(c)). This approach fully confirms the results shown in figure~\ref{fig:2D-FFT}, namely that each separated mode is indeed associated with the spiral components traveling upward and downwards without any changes in the frequency, but with changes in their wavenumbers. This implies that a linear superposition of the two spirals should explain some part of the amplitude modulation. As an extension of the analysis, we also examined the Radon transform on a more complex case involving a taller cavity with $H = \SI{2800}{\milli\metre}$ (four times taller than the configuration considered here). In this setup, the flow exhibits more complicated spiral patterns, but all the conclusions we can obtain from those results are consistent with those presented here, i.e., we observe similar regions being activated/deactivated depending on whether the spiral is moving upwards or downwards in a given time. These additional results are presented in Appendix~\ref{appendix:LargerCavity}.

\begin{figure}
    \centering
    \renewcommand{\arraystretch}{1.2}
    \begin{tabular}{c|c}

    \textbf{Upward traveling spiral} & \textbf{Downward traveling spiral} \\
    
        \begin{minipage}[t]{0.45\linewidth}
            \centering
            \includegraphics[width=\linewidth]{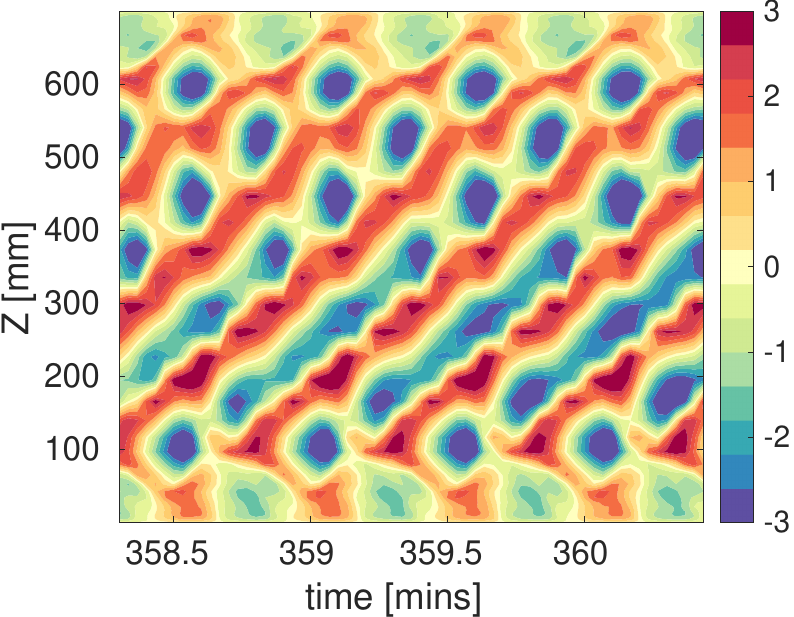}
            \caption*{(a)~$u_\phi-\overline{u_\phi}^t$ (full signal)}
        \end{minipage}
        &
        \begin{minipage}[t]{0.45\linewidth}
            \centering
            \includegraphics[width=\linewidth]{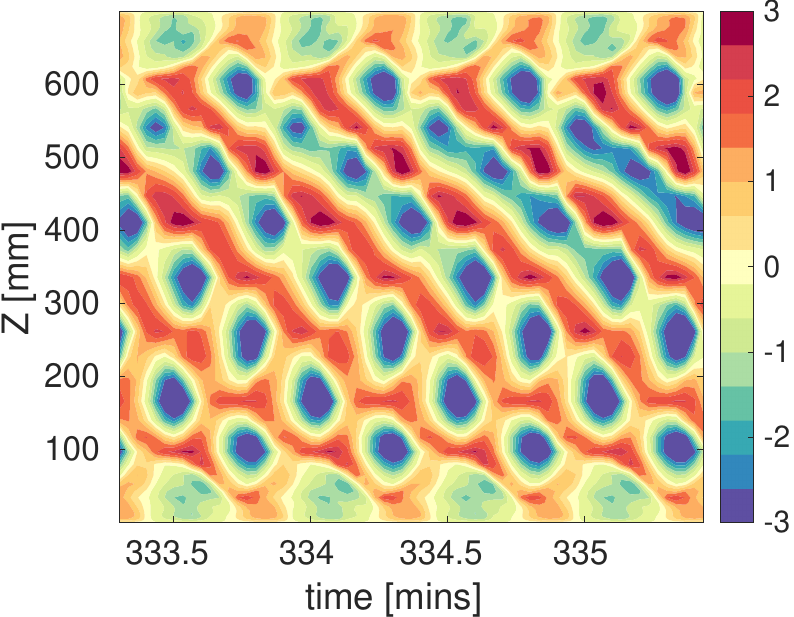}
            \caption*{(b)~$u_\phi-\overline{u_\phi}^t$ (full signal)}
        \end{minipage}
        \\[2ex]
        
        \begin{minipage}[t]{0.45\linewidth}
            \centering
            \includegraphics[width=\linewidth]{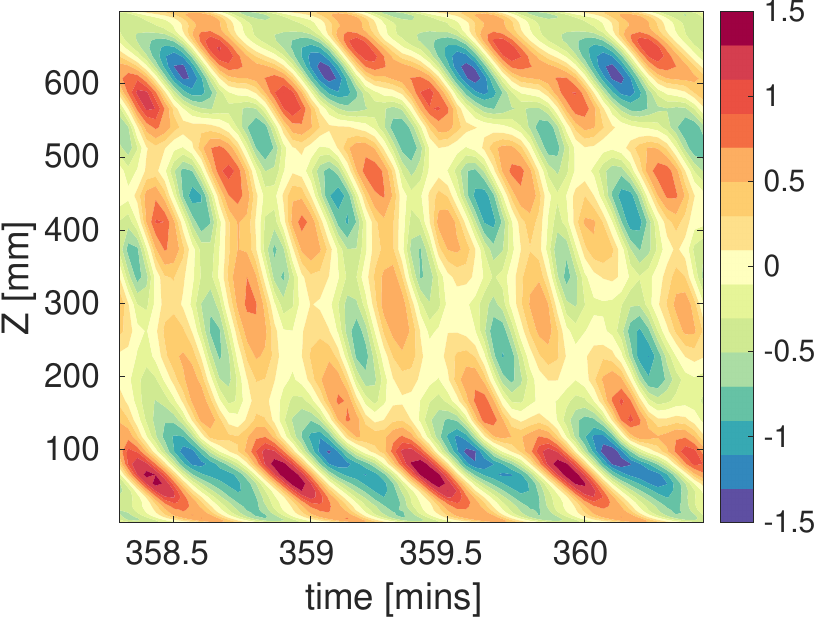}
\caption*{(c)~Downward signal of the Upward traveling spiral}
        \end{minipage}
        &
        \begin{minipage}[t]{0.45\linewidth}
            \centering
            \includegraphics[width=\linewidth]{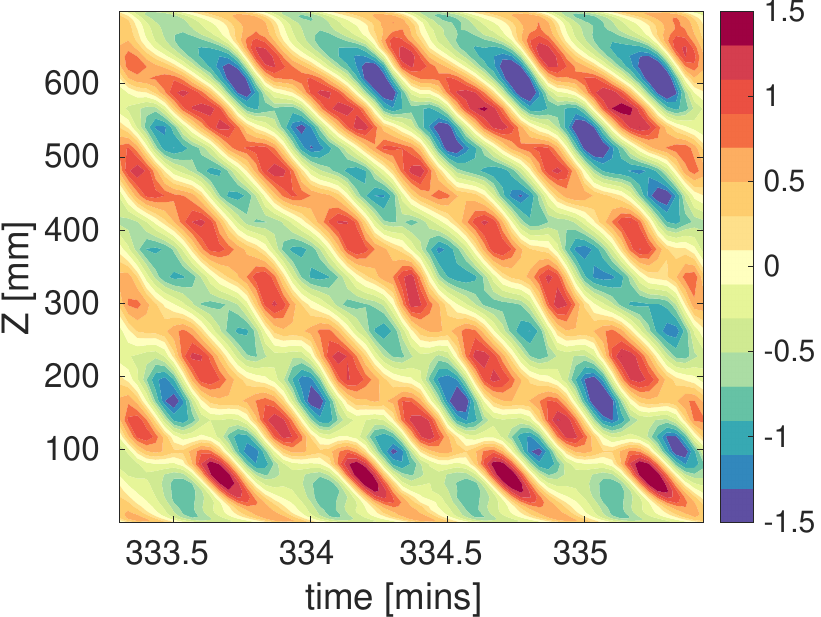}
            \caption*{(d)~Downward signal of the downward traveling spiral}
        \end{minipage}
        \\[2ex]
        
        \begin{minipage}[t]{0.45\linewidth}
            \centering
            \includegraphics[width=\linewidth]{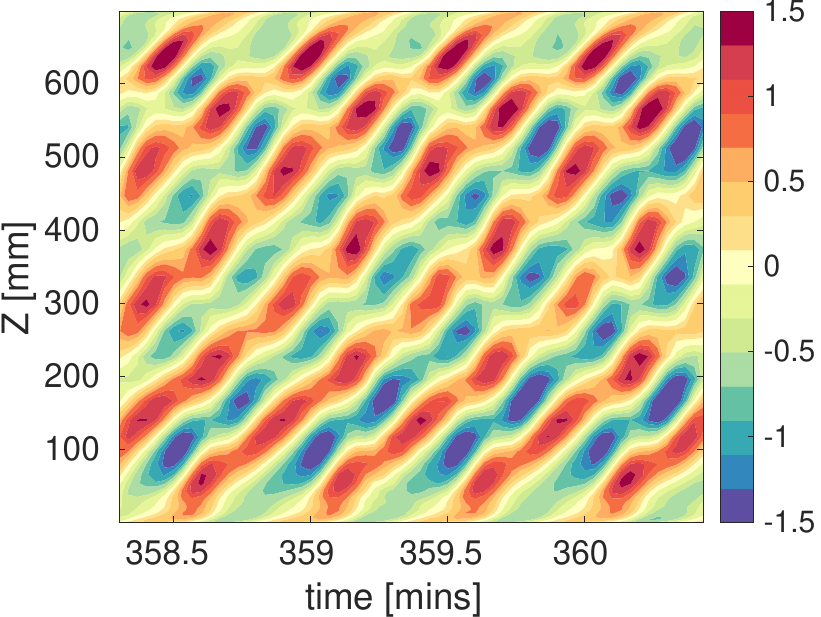}
            \caption*{(e)~Upward signal of the Upward traveling spiral}
        \end{minipage}
        &
        \begin{minipage}[t]{0.45\linewidth}
            \centering
            \includegraphics[width=\linewidth]{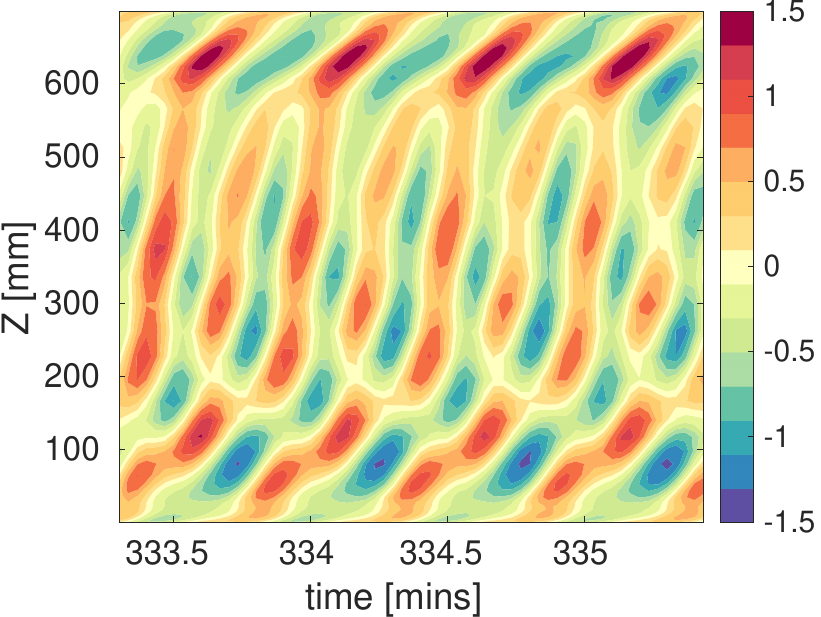}
            \caption*{(f)~Upward signal of the Downward traveling spiral}
        \end{minipage}

    \end{tabular}
    \caption{Separation of upward and downward axial traveling components in $u_\phi$ space-time diagram using the Radon Transform. Figures~(a),(c),(e), on the left-hand side, show time intervals when the spiral is traveling upwards. Figures~(b),(d),(f), on the right-hand side, show time intervals when the spiral is traveling downwards. Figures~(a),(b) on top show the full space-time diagram minus the mean flow (computed using the full time signal). Figures~(c)--(f) show the separated upward and downward traveling components obtained using the Radon Transform (applied to the full signal, but it filters out the mean flow when it is applied). Simulation performed with $Re=400$, $\mu = 0.35$, $\Delta T/\Delta z \approx \SI{5.71}{\kelvin \per \metre}$, and $H=\SI{700}{\milli \metre}$. \label{fig:Radom_Re400}}
\end{figure}

\begin{figure}
	\begin{center}
		\begin{minipage}[t]{0.42\linewidth}
			\centering
			\includegraphics[width=1\linewidth]{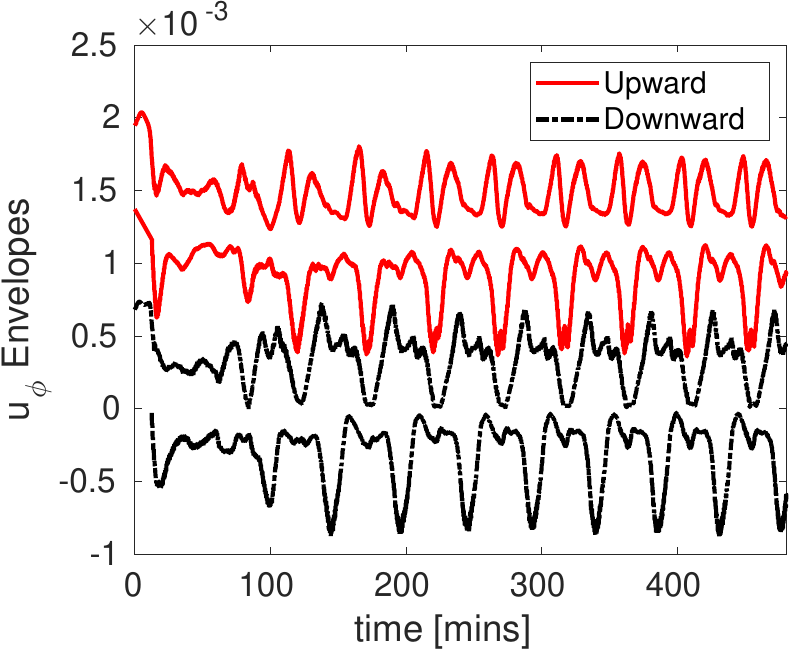}
			\caption*{{(a)}} 
			\vspace{2ex}
		\end{minipage}
		\hspace{.1in}
		\begin{minipage}[t]{0.42\linewidth}
			\centering
			\includegraphics[width=1\linewidth]{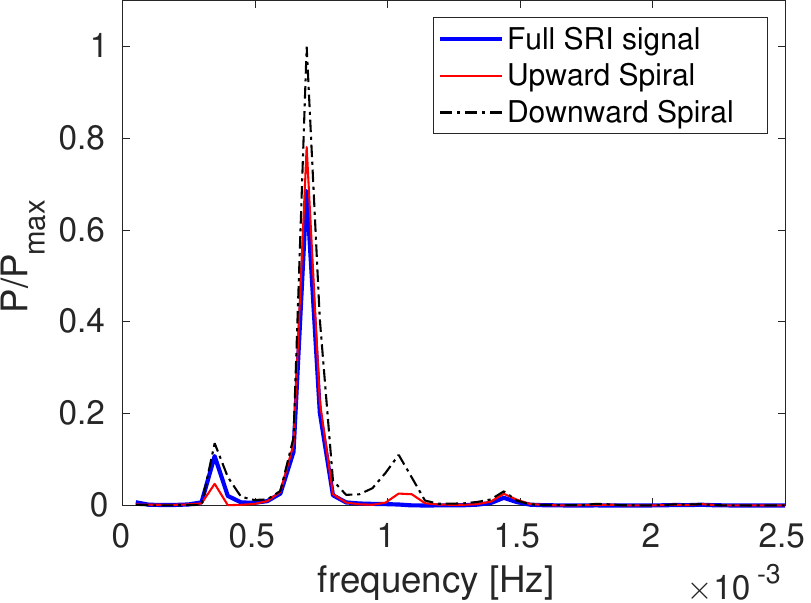}
			\caption*{{(b)}} 
			\vspace{2ex}
		\end{minipage} 
		\caption{(a)~$u_\phi^\prime$ upward and downward spiral components amplitude modulations. The upward traveling time series was arbitrarily dislocated in the vertical axis for better visualization (originally, both time series were on top of each other). (b)~Power spectra obtained from the amplitude envelopes of the velocity time series, and of their separated upward and downward components, obtained using the Randon transform to separate the signals. The time series is obtained from numerical simulations with $Re=400$, $\mu=0.35$ and $\Delta T/\Delta z \approx \SI{5.71}{\kelvin \per \metre}$ at mid-gap position ($r \approx r_{in}+d/2$) and mid height position ($z \approx H/2$). 
		\label{fig:UpDown_Envelope} }
	\end{center}
\end{figure}

Figure~\ref{fig:UpDown_Envelope}(a) shows the time series envelope of separated upward and downward spiral components. 
  Note that this is not the full SRI velocity, which presents faster oscillations (see~\cite{Meletti2020GAFD}), but just the envelope capturing the velocity amplitude modulations. The upward traveling time series was arbitrarily dislocated in the vertical axis for better visualization (originally, both time series were on top of each other). Note that, also here, when the amplitude of the downward spiral is enhanced, the upward spiral amplitude becomes smaller (and vice-versa), showing a phase shift of the up-downward components beating. When the power spectrum of $u_\phi$ amplitude envelope is compared to the separated upward and downward components obtained by using the RT in figure~\ref{fig:UpDown_Envelope}(b), the same (low) frequencies are observed.


	
%
Note that we can observe harmonics in the envelope spectra presented in fig.~\ref{fig:UpDown_Envelope}.(b), and also on each of the separated up~and~downward spiral components, which suggests a weak non-linear interaction, and not simply a linear interaction of two waves traveling with different frequencies. 
%
Note also that the amplitudes of each separated spiral component in the power spectra differ from those obtained from the full velocity signal. A similar behavior is obtained from both $u_\phi$ and $u_r$ time series, i.e., with peaks corresponding to the same frequencies in \si{\hertz}, but with different amplitudes. 
%
%
%
But most importantly, we would like to highlight that, when the Radon filtering is applied, we observe a clear 
phase shift in the modulations of each separated component. This phase shift indicates that the harmonics observed in the spectra are not simply modulated in isolation; rather, they suggest a non-linear interaction, causing the modulations. The fact that these modulations cannot be fully explained by linear interactions alone implies that a more complex dynamic is at play, where the interaction between the spirals and the mean flow results in the alternating strengthening and weakening of the SRI spiral components. This interaction, in turn, leads to the observed alternation between upward and downward propagating modes, each with varying amplitudes. In the following sections, we will investigate how these two individually modulated components could lead to the pattern formations we observed.

%% file: content/ToyModel.tex
\section{ Toy model: Wave-like spiral propagation} \label{sec:ToyModel}


In this section, we introduce a toy model consisting of two waves traveling in opposite axial directions to demonstrate how the linear superposition of the SRI spirals can lead to pattern changes. This approach is inspired by the separation of the spirals using the Radon Transform (RT) technique. We observed in the previous sections that each spiral (traveling upward and downward) is individually modulated and phase-shifted relative to one another. The reasons behind the modulation of each spiral will be discussed later in this paper. For now, we assume that the spirals behave as two individual plane waves, described by the following equations:

\begin{equation}\label{eq:ToyModel}
    \begin{aligned}
     &   \text{wave}_1 = A_1 cos \left( \left( m_1x + l_1y + k_1 z \right) - \omega t \right), &  \\
     &   \text{wave}_2 = A_2 cos \left( \left( m_2x + l_2y + k_2 z \right) - \omega t \right), & \\
     &   u_{toy} = \text{wave}_1 +  \text{wave}_2 , & \\
    \end{aligned}
\end{equation}

\noindent with $0 \leq x,y,z \leq 2 \pi$, and amplitudes $A_1$ and $A_2$. The values of $0<x,y,z<2\pi$ result from normalizing the cavity height (e.g., $\SI{0}{\milli\metre} < z < \SI{700}{\milli\metre}$).  The amplitude of each plane wave is modulated, and they must be out of phase to achieve the inclined spirals with constructive and destructive interference while they propagate. In the toy model,  sinusoidal amplitude modulations $A_1$ and $A_2$ are considered out of phase with an angle $\theta$, written as
\begin{equation}\label{eq:ToyModelAmplitude}
    \begin{aligned}
    &    A_1 = A sin(\omega_A t) , &  \\
    &    A_2 = A sin(\omega_A t+\theta) , & \\
    \end{aligned}
\end{equation}
\noindent where $A$ is a given real value, and $\omega_A << \omega$ is the amplitude modulation of each wave, here considered to be the same for $A_1$ and $A_2$.

\begin{figure}
    \centering
            \includegraphics[scale=0.4]{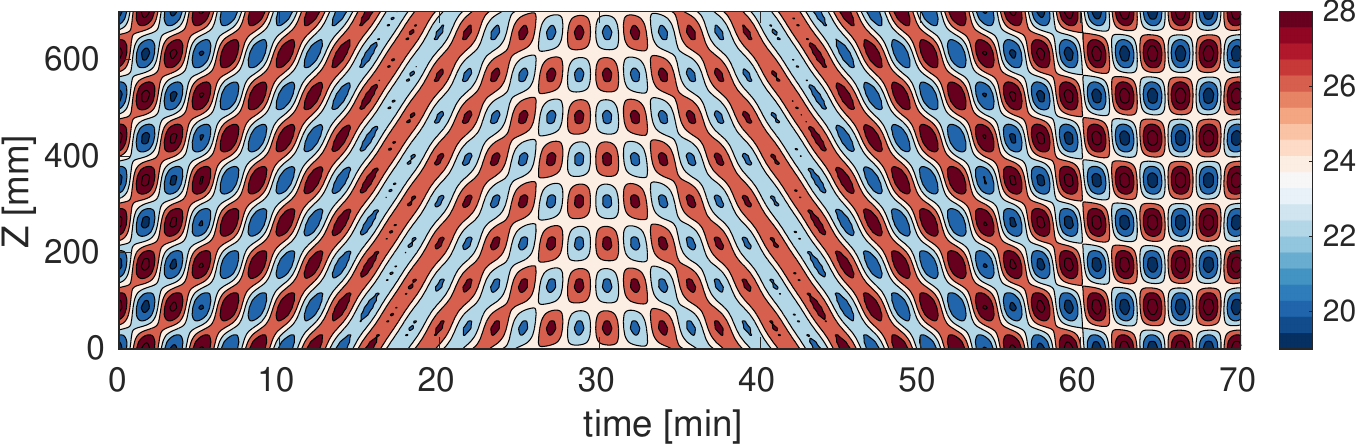}

    \caption{$u_\phi$ space-time diagram of the toy model composed of 2 plane waves with sinusoidal amplitude modulations with $\omega_A=7\times10^{-4}$, out of phase an angle $\theta=\pi/3$, and traveling in opposite axial directions with wavenumbers of $\text{wave}_1$ and $\text{wave}_2$ respectively \mbox{$(m_1,l_1,k_1)= (1,1,4)$} and \mbox{$(m_2,l_2,k_2)=(1,1,-4)$}. The frequency $\omega = 0.03$, and the maximum amplitude of each wave is $A=10$.
    \label{fig:Hov_Toymodel} }
\end{figure}

\begin{figure}
    \begin{center}
    \renewcommand{\arraystretch}{1.2}
    \begin{tabular}{c|c}
        \textbf{Simulation} & \textbf{Toy model} \\
        
        \begin{minipage}[t]{0.4\linewidth}
            \centering
            \includegraphics[width=0.95\linewidth]{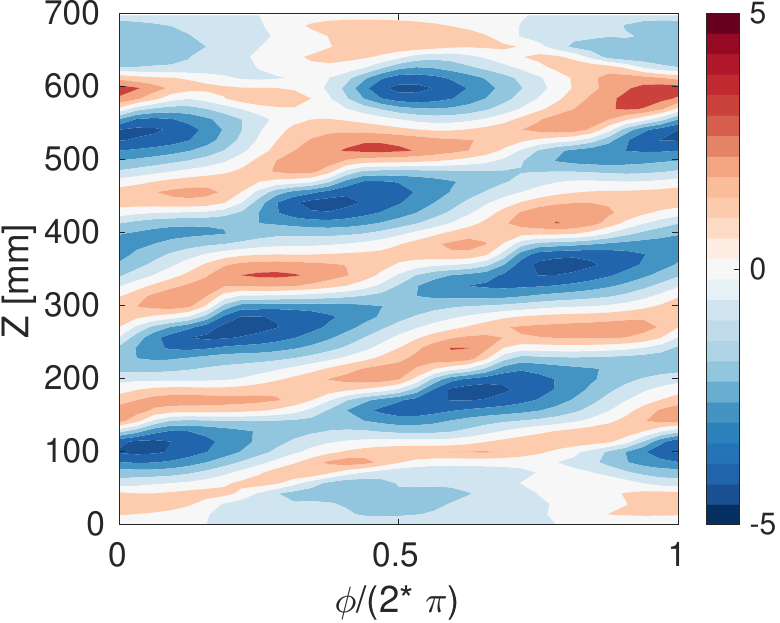}
            \caption*{(a)~Simulation downward}
        \end{minipage}
        &
        \begin{minipage}[t]{0.4\linewidth}
            \centering
            \includegraphics[width=0.95\linewidth]{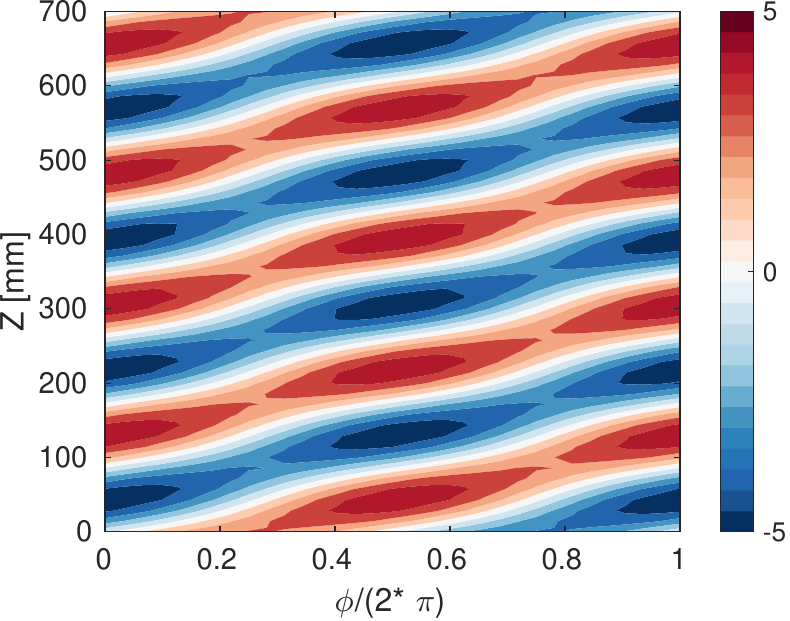}
            \caption*{(d)~Toy model downward}
        \end{minipage}
        \\

        \begin{minipage}[t]{0.4\linewidth}
            \centering
            \includegraphics[width=0.95\linewidth]{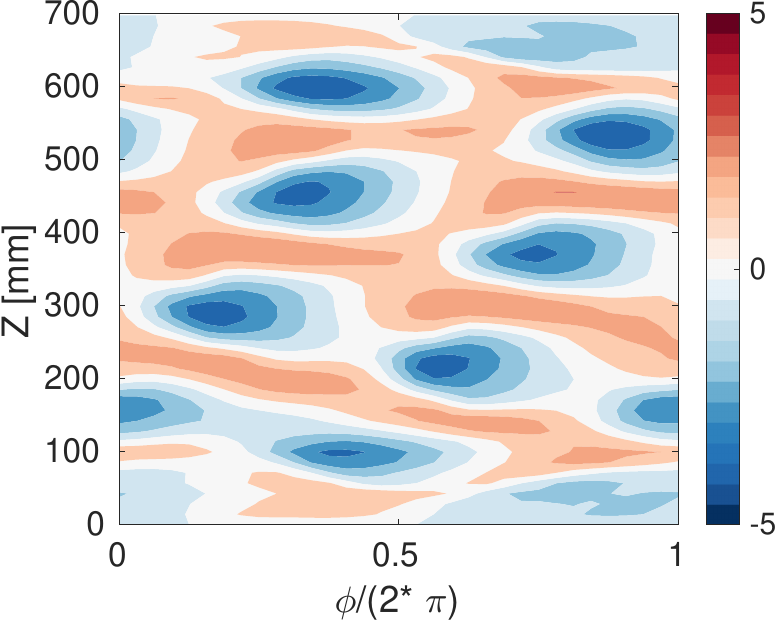}
            \caption*{(b)~Simulation transition}
        \end{minipage}
        &
        \begin{minipage}[t]{0.4\linewidth}
            \centering
            \includegraphics[width=0.95\linewidth]{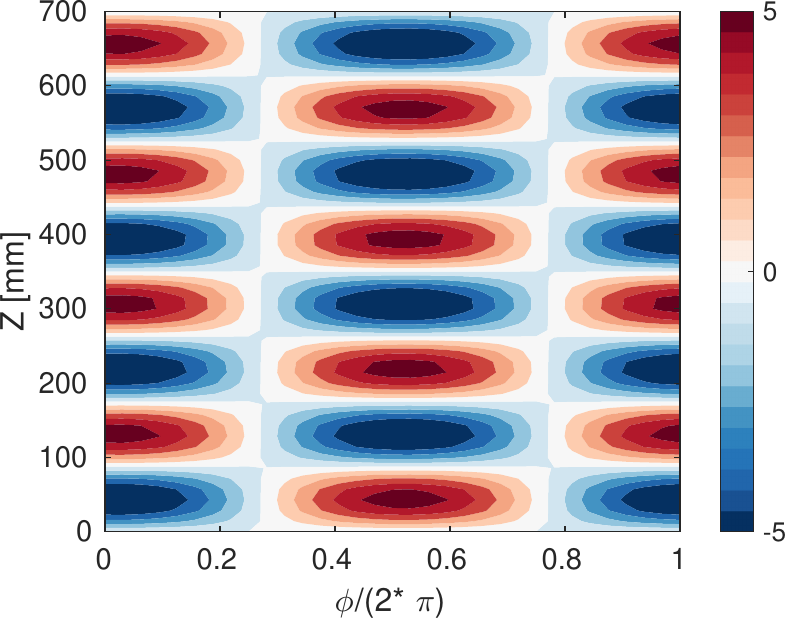}
            \caption*{(e)~Toy model transition}
        \end{minipage}
        \\

        \begin{minipage}[t]{0.4\linewidth}
            \centering
            \includegraphics[width=0.95\linewidth]{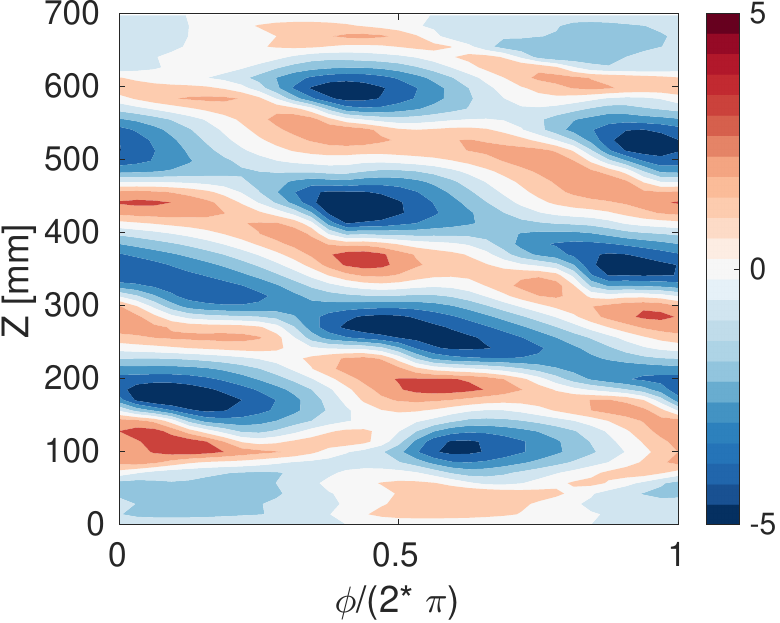}
            \caption*{(c)~Simulation upward}
        \end{minipage}
        &
        \begin{minipage}[t]{0.4\linewidth}
            \centering
            \includegraphics[width=0.95\linewidth]{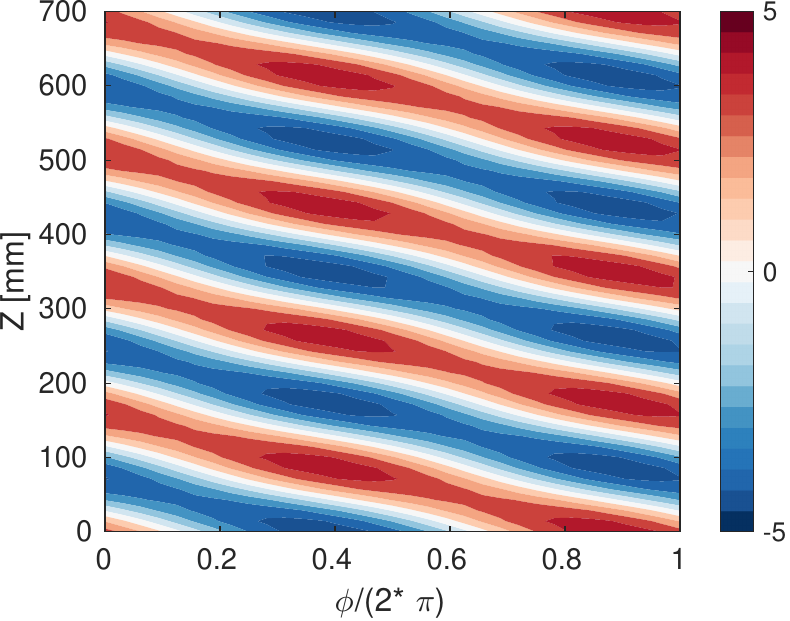}
            \caption*{(f)~Toy model upward}
        \end{minipage}
    \end{tabular}
        


    
    \caption{Snapshots with different spiral patterns in the $\phi$-$z$ cross-section comparing $u_\phi^\prime = u_\phi -\overline{u_\phi}$ obtained from numerical simulations fixed at a radial position $r\approx r_{in}+d/3$ (a,b,c) and the toy model (d,e,f). Figures~(a),(d) on top show moments when the spirals are traveling downwards; (b),(e) show the transition; and (c),(f) show spirals traveling upwards. The simulations were performed with $Re=400$, $\mu=0.35$, and $\Delta T/\Delta z \approx \SI{5.71}{\kelvin\per\metre}$. The toy model consists of two plane waves with frequencies $\omega=0.001$ and wavenumbers $(m_1,l_1,k_1)= (1,1,4)$ and $(m_2,l_2,k_2)=(1,1,-4)$, wave amplitude $A=3$~mm/s, and modulation frequency $\omega_A=0.01$ with $\theta=\pi/2$ phase difference. \label{fig:Toymodel_snapshots}}
    \end{center}
\end{figure}


Figure~\ref{fig:Hov_Toymodel} shows the space-time diagram obtained with this toy model, with wavenumbers in the azimuthal, radial and axial directions ($m=1,l=1,k=4$) and ($m=1,l=1,k=-4$), similar to the ones previously observed in our simulations (with $Re=400$, $\mu=0.35$,$\Delta T/\Delta z \approx \SI{5.71}{\kelvin\per\metre}$, and $H=\SI{700}{\milli\metre}$). It is possible to see that the linear superposition of both upward and downward waves, each individually modulated and traveling out of phase, could lead to the final spiral pattern transitions observed in the previous sections. The amplitude modulations of the waves presented in figure~\ref{fig:Hov_Toymodel} are out of phase with an angle $\theta=\pi/3$, but other different phase shifts (and different wavenumbers) produce similar pattern changes.

The toy model and the numerical simulation also show good qualitative agreement when we compare snapshots, i.e., looking at the space structures at given times while the spirals are traveling upward, downward, and during the transition. This comparison can be seen in figure~\ref{fig:Toymodel_snapshots}, where we can see that the results are similar, except for the fact that the spirals in the simulations are confined in a slightly smaller region, due to Ekman effects. The good qualitative agreement of this simplified toy model with the numerical simulations, as well as the possibility of reproducing the spiral pattern changes previously investigated using similar wavenumbers and frequencies obtained from numerical simulations and experimental measurements, shows that this linear superposition of the spirals can drive the spiral pattern changes. In this case, each up~and~downward component should be interacting with the mean flow that, at times, provides more energy to the upward traveling spiral, and at other times, provides more energy to the downward component. The reason why each individual spiral is modulated will be interpreted as a QBO-like mechanism in the following section, inspired by the fact that the spiral propagation direction affects the mean flow structure (see figure~\ref{fig:MeanPerRegion}). Therefore, each spiral component must be interacting with the mean flow to achieve its individual (phase-shifted) modulation, so that the linear superposition of these two modulated components will lead to the reversal of the spiral direction as presented here in this toy model.

%% file: content/AxialMeanFlow.tex
\section{Axial mean flow interpretation considering inertial wave interactions}\label{sec:AxialMeanFlow}
\begin{figure}
\centering
\includegraphics[width=0.4\linewidth]{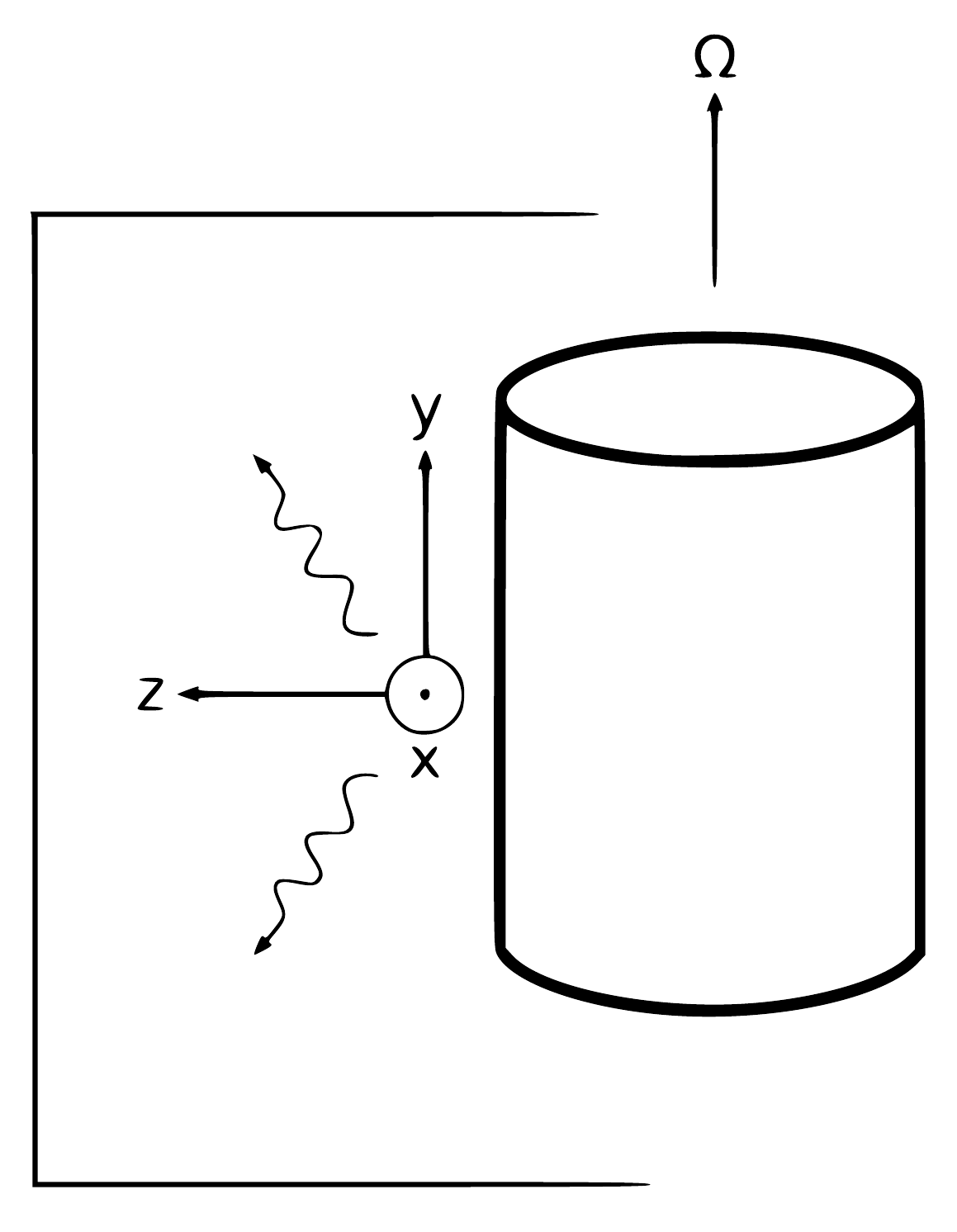}
\caption{Local Cartesian coordinate system used for (\ref{iw1})-(\ref{iw3}). Waves propagating in the $z$-$y$-plane are sketched. Note that in such a ``planetary model'' the rotation vector and the gravity vector are perpendicular to each other. In contrast, these vectors would be parallel and in the opposite direction in the SRI-cylinder. In our model, we neglected the explicit gravity term for simplicity.} \label{fig:Coord} 
\end{figure}

Based on previous observations of mean flow variations associated with the oscillatory spiral behavior, and given that the toy model introduced in the previous section successfully reproduces these pattern changes, we will now interpret the low-frequency amplitude modulation of the SRI as a Quasi-Biennial Oscillation (QBO)-like phenomenon. Furthermore, we note by identifying one inner cylinder rotation with one SRI-day that the mean flow variations take about 2.6 SRI-years, when we compare the first low-frequency peak observed in figure~\ref{fig:UpDown_Envelope}.(b) with the inner cylinder's rotation (from which the Reynolds number is computed). The velocity amplitude modulations we observe in figure~\ref{fig:UpDown_Envelope}.(a) are linked to the traveling spiral direction (fig.~\ref{fig:Pattern_modulation}) in a weakly non-linear manner, given that we see harmonics in the envelope spectra, indicating an interaction between the base flow and the instability. To consider these modulations as a QBO occurring in a stably stratified annular flow setup, our approach follows the ideas presented in \citet{plumb1977interaction, seelig2015can, renaud2020holton}, but focuses on inertial waves due to the weak stratification in the SRI simulation. Instead of considering the classical non-rotating but stratified QBO scenario, where the base flow in the azimuthal direction produces a QBO in the azimuthal-radial plane, we consider the base flow in the axial direction generating the QBO in the axial-radial plane. In other words, we will show here how the wave interactions in the axial-radial plane can drive an oscillating axial mean flow. Since, for the flow discussed here, the ratio between the Coriolis parameter $f=2 \Omega_{in}$ and the buoyancy frequency $N$ is larger than 1 ($f/N=3$), we focus here on inertial waves and not on internal gravity waves as in the model by \citet{plumb1977interaction}. However, we will see that the equations are analogous to equations for the atmospheric QBO in the equatorial stratosphere, driven by internal gravity waves. The analysis is local Cartesian, as presented in fig.~\ref{fig:Coord}; that is, we neglect curvature and any azimuthal variation. In fact, for the SRI, the azimuthal wave number is $m=1$, but taking a local viewpoint, the variation in the azimuthal direction is small. For further simplification, the diffusion terms in the \citet{plumb1977interaction} model are replaced by viscous drag terms with a drag coefficient $\varsigma$. Then the momentum equations in a frame co-rotating with $\Omega_{in}$ read
\begin{eqnarray}
\partial_t u + v\partial_y u + w\partial_z u &=& -2\Omega_{in} w -\varsigma u, \label{iw1}\\
\partial_tv+v\partial_yv+w\partial_zv &=& -\frac{1}{\rho_0} \partial_y p -\varsigma v, \label{iw2}\\
\partial_tw+v\partial_yw+w\partial_zw &=& -\frac{1}{\rho_0}\partial_z p + 2\Omega_{in} u - \varsigma w,\label{iw3}
\end{eqnarray}
where $(x,y,z)$ are the azimuthal, axial, and radial directions, respectively, and $(u,v,w)$ are the azimuthal, axial, and radial velocity components (see Fig. \ref{fig:Coord}). Here, $p$ is the generalized pressure that includes centrifugal effects. 

Defining the streamfunction  $\psi=\psi(y,z,t)$  and a zonal momentum
$\tilde u$ as
\begin{eqnarray}
v=-\partial_z \psi, \quad w=\partial_y\psi, \quad \tilde{u}=2\Omega_{in} u,
\end{eqnarray}
\noindent this system can be reduced to
 \begin{eqnarray}
\partial_t\nabla^2\psi+J(\psi,\nabla^2\psi) & = & \partial_y\tilde{u}-\varsigma\nabla^2\psi  \label{in1}\\
\partial_t\tilde{u}+J(\psi,\tilde{u}) & = & -4\Omega^2_{in}\partial_y\psi-\varsigma\tilde{u} . \label{in2}
\end{eqnarray}
By introducing the streamfunction in the azimuthal-radial plane, 
\begin{eqnarray}
\psi = \psi(x,z,t), \quad u = -\partial_z\psi, \quad w = \partial_x \psi, 
\end{eqnarray}
\noindent and with a gravity vector pointing in the negative $z$-direction, \cite{plumb1977interaction} derived the following equation for the case of internal gravity waves
\begin{eqnarray}
\partial_t\nabla^2\psi+J(\psi,\nabla^2\psi) &=& \partial_xb+\nu\nabla^4\psi  \label{gr1}\\
\partial_{t}b+J(\psi,b) & = & -N^{2}\partial_{x}\psi-\mu b,  \label{gr2}
\end{eqnarray}
where $b=-g \rho'/\rho_0$ is buoyancy, $\rho'$ and $\rho_0$ is the density perturbation and the mean density, $\nu$ is the kinematic viscosity and $N^2=-(g/\rho_0) d \bar \rho/dz$ is the square of the buoyancy frequency, where $\bar \rho$ is the background density with a linear dependency on $z$. Here, there is no rotation, and in contrast to the previous case, the waves propagate in the azimuthal-radial plane (the $x$-$z$-plane). Except the linear friction term in (\ref{in1}), the equations (\ref{gr1}), (\ref{gr2}) and (\ref{in1}),
(\ref{in2}) are mathematically isomorphic. 
Friction terms are also used, e.g., in simple models for vorticity Ekman pumping \citep{pedlosky1987parsons}.

To derive a ``QBO model'' from Eqs.~(\ref{in1})~and~(\ref{in2}), we can proceed in the same way as for the gravity waves.
First, a mean state $\bar \psi$ is defined, the streamfunction is written as $\psi(y,z,t) = \bar \psi(z)+\epsilon \psi'(y,z,t)$, and the equations are linearized about the mean streamfunction $\bar \psi$, where $\psi'(y,z,t)$ is the fluctuation around the mean value, and $\epsilon$ is a small perturbation parameter, $\epsilon \ll 1$. The perturbation is expanded as
$\psi'(y,z,t)=\phi(z) \exp(\mathrm{i}l(y-ct))$, where $l$ is the axial wave number and $c$ the axial phase speed.
We obtain a linear equation for $\phi(z)$ and its solution
gives the perturbation velocities $v'$ and $w'$. Then (\ref{iw2}) is averaged over a wavelength in the {\em axial} direction to obtain
\begin{equation}
\partial_{t} \bar v +\varsigma \bar v=-\epsilon^2 \partial_{z} (\overline{v'w'}),\label{eq:vbar}
\end{equation}
where $\bar v=-\partial_z \bar \psi$ is the slowly varying mean flow in the $y$-direction.

\cite{plumb1977interaction} found for the atmospheric QBO case
\begin{equation}
\partial_{t} \bar u -\nu \partial_{zz} \bar u = -\epsilon^2 \partial_{z} (\overline{u'w'}),\label{eq:ubar}
\end{equation}
i.e., a slowly varying mean flow in the $x$-direction driven by waves in the $x$-$z$-plane. Again, we see a strong analogy between (\ref{eq:vbar}) and (\ref{eq:ubar}).
\cite{plumb1977interaction} was able to ``parameterize'' the wave momentum flux as
\begin{equation}
    (\overline{u'w'}) \sim \exp \left( -s \int_0^z \frac{N \mu}{k (\bar u-c)^2}dz' \right)
\end{equation}
where $s$ is the sign of the $z$-component of the group velocity, $k$ is the wavenumber in the $x$-direction and $c=\omega/k$ is the phase velocity, where $\omega$ is the wave frequency. In analogy, we can write for the inertial wave case
\begin{equation}
    (\overline{v'w'}) \sim \exp \left( -s \int_0^z \frac{2 \Omega \zeta}{l (\bar v-c)^2}dz' \right).
\end{equation}
Considering two waves propagating towards the positive and negative $y$-direction, as proposed in the previous section, the mean flow equation can be written in the universal non-dimensional form
\begin{equation} \label{QBOmodel}
    \partial_{t} \bar v + Re^{-1} \bar v=- \epsilon^2 \frac{\partial}{\partial z}  \left( \exp \left( - \int_0^z \frac{1}{(\bar v-1)^2} dz' \right) - \exp \left( - \int_0^z \frac{1}{(\bar v+1)^2} dz' \right)\right),
\end{equation}
where $Re^{-1}=L \varsigma/V$ for inertial waves and $Re^{-1}=\nu/(LU) \partial_{zz}$ for internal gravity waves \citep{RenaudQBO}.
\begin{figure}
\centering
\includegraphics[width=1.0\linewidth]{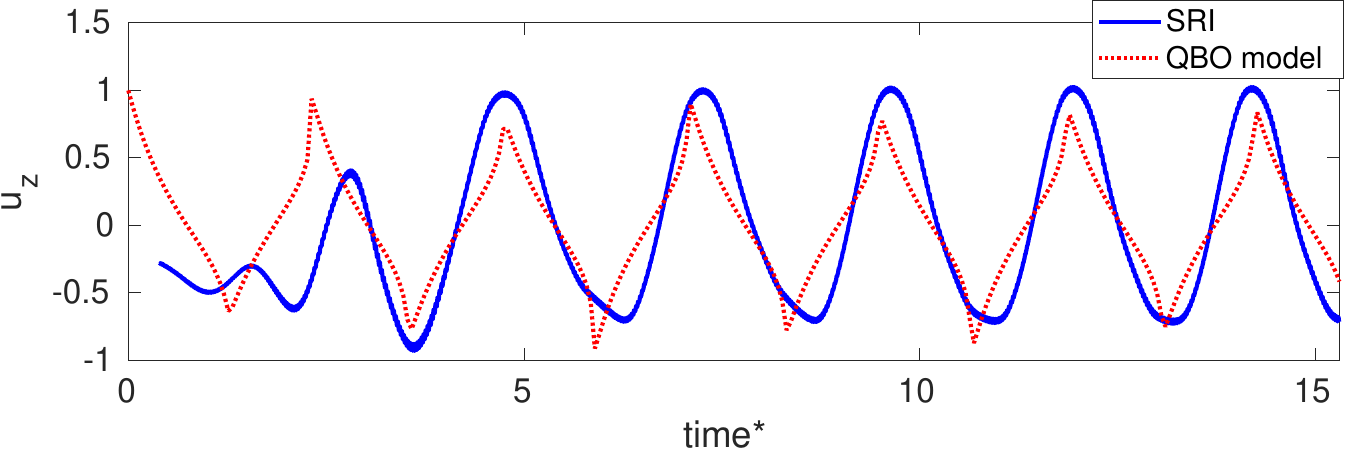}
\caption{Time-series of the axial flow $\bar v(z)$ comparing the moving average of the DNS SRI simulation results $-$~at a radial position at radial position $r=r \approx r_{in}+d/3$ and mid height position (computed from Eq.~\ref{eq:Numeric_Eqn})~$-$ with the QBO model (eq.~(\ref{QBOmodel})) at the position $z=1/3$, considering the coordinates system presented in Fig.~\ref{fig:Coord}. The DNS results were filtered with a 15~minutes moving average to highlight the low-frequency components. The \textit{time*} of the numerical simulation results was normalized considering the period of $T\approx2.6$ SRI-years observed for the low-frequency peaks (compared to the inner cylinder frequency).}
%
\label{QBO_Hov} 
\end{figure}

We solved (\ref{QBOmodel}) numerically using the 4th-order Runge-Kutta scheme for the time derivative and the trapezoidal rule for the integrals. The result is shown as a red-dotted line in Fig.~\ref{QBO_Hov}, where we plotted the axial velocity at the position z=1/3, ${\overline v}(t,z=1/3)=u_z(t)$, as a function of time. 
It can be seen that over the whole time considered, the direction of the axial flow oscillates. The largest amplitudes can be found moving towards $z=0$ in the coordinates system presented in Fig.~\ref{fig:Coord}, which corresponds to the area near $r=r_{in}$ in the DNS results, but similar oscillatory behavior could be seen practically at any chosen position (then we chose a region near $z=1/3$, that corresponds to $r=r\approx r_{in}+d/3$ in the DNS results). The amplitude weakens away from the inner boundary. Note that the mean flow converges to a steady state for a single wave with a narrow boundary layer at the inner cylinder wall and a constant flow above this layer. Adding a second wave, the mean flow becomes unstable and starts to oscillate. More details can be found in \citet{plumb1977interaction} and \citet{RenaudQBO}. Note that the mechanism described here has been reproduced in a laboratory experiment with internal gravity waves by \citet{Plumb_McEwan}. In Fig.~\ref{QBO_Hov}, we also show the profiles of $u_z$ obtained using the full SRI simulations as a full blue line, for a qualitative comparison. These results were filtered with a 15~minutes moving average to highlight the low-frequency variations of the mean flow. This time window was chosen arbitrarily. Other periods were also tried for the windowing, and the qualitative features observed do not change considering our purposes here, i.e, using it as a low-pass filter to highlight the low frequency features. The time in the y-axis was normalized so that the period of the low-frequency oscillation would take approximately $T\approx2.6$ SRI-years, as observed in our data. We highlight here that the general features on the two plots are not very different, with similar QBO-like oscillations observed, and even with boundary effects influencing the flow until a height at the lower boundary in both results.  The real SRI configuration modeled with the DNS code is more complicated than what is described by the simplified QBO-like model we presented here, since, for example, the full top and bottom boundary layer effects present in the experimental cylinder are not taken into account by the model. Still, it has been shown in \citet{meletti2023parameter}, in axial periodic simulations, that the axial pattern changes do not need boundaries to occur, although they have some influence. Therefore, we do not attempt to justify the realism of the inertial wave–mean flow model on modeling the SRI spiral pattern changes. Rather, we use this simplified model to show how it produces similar results to support our interpretation of the spiral propagation changes in the axial direction as QBO-like features. We are concerned, here, with illustrating how the qualitative features of the observed low-frequency oscillations in the SRI can be interpreted as resulting from mean-flow/instability interactions.

%% file: content/conclusion.tex
\section{Conclusions} \label{sec:conclusion}

This study explores the interactions between axial modes and spiral components in a stratified, rotating flow that develops Strato-Rotational Instability (SRI). 
 The findings demonstrate that the SRI engenders complex oscillatory dynamics, previously observed in \cite{Meletti2020GAFD}, significantly alter the mean flow. The oscillations are characterized by the selective activation of distinct axial wavenumbers, corresponding to upward and downward propagating spiral modes that alternate in dominance.
 
The application of Radon Transform (RT) enabled a clear separation of the upward and downward spiral components, revealing their interaction with the mean flow. This interaction manifests as amplitude modulations, which have been robustly captured in both numerical simulations and experimental data \citep{Meletti2020GAFD,meletti2023parameter,lopez2022stratified,Riedinger2010}. We observed here the presence of out-of-phase modulated harmonic waves associated with specific azimuthal wavenumbers, further indicating that the dynamics are governed by non-linear wave-mean flow interactions. Inspired by these RT findings, a simplified toy model was developed to interpret the physics underlying the spiral pattern changes associated with amplitude modulations. This model proposes that the observed phenomena can be interpreted as two individual wave-like spirals propagating in opposite directions along the axial axis, added in a simple linear superposition. The RT findings also suggested that these spirals should be individually modulated and that these modulations should be out of phase. This was incorporated into the toy model. By linearly superimposing these two modulated wave-like traveling spirals, the toy model successfully reproduced the observed spiral pattern changes, providing an interpretation of the underlying physics observed.

However, the important question of the origin of the individual modulation of each spiral still had to be addressed. The observed base flow/instability interaction, indicated by changes in the mean flow, inspired the comparison of the individual modulations to a quasi-biennial oscillation (QBO)-like phenomenon. To model this, we employed a simplified linearized QBO-like dynamics approach, following the models proposed by \citet{plumb1977interaction,seelig2015can,RenaudQBO} to derive a QBO-like model in the axial direction, explaining the mean-flow/instability mechanisms. This model successfully replicates the qualitative features observed in the axial velocity, confirming that the QBO-like behavior that arises from the mean-flow/instability interaction can explain the individual spiral modulations previously introduced in the wave-like toy-model.

%% file: content/Appendices.tex
\newpage
\appendix

\section{Radon Transform} \label{appendix:Radon}

The Radon transform $R(r,\phi)$ consists of a Fourier-like technique developed by \citet{radon1917bestimmung}. This technique transforms a function defined on a given plane $\eta(z,t)$ into a line domain. These lines are inside the original 2-d~space, with the values of a particular line being equal to the line integral of the original function (over that projected line). Therefore, the Radon transform consists on an angular projection given by:

\begin{equation}\label{eq:appendix_Radon}
    R(r,\phi) = \oiint \eta(z,t)\delta \left(z cos\phi + t sin \phi - r \right)dz dt,
\end{equation}

\noindent where $\delta$ the Dirac delta function. $r = z cos \phi + z sin \phi$ and $\phi$ are respectively the radius and angle, in polar coordinates, that define the line where the 2-D~space will be projected. $\phi$ can vary from $0$ to $\pi$. The use of the Dirac delta function forces the integration of $\eta(x,t)$  along the line on which the plane will be projected. If  we consider a  two-dimensional  spatiotemporal  wave  signal $\eta(z,t)$, traveling in the $z$ direction,  the  angle $\phi$ can  be converted into a wave drift velocity $c$ through the transformation \citep{almar2014use}

\begin{equation}\label{eq:appendix_RadonPhaseSpeed}
    c = tan(\phi) \frac{dz}{dt},
\end{equation}

\noindent where $dz$ and $dt$ are respectively the spatial and temporal resolution.  If the $\eta(z,t)$  signal contains multiple  waves,  multiple  local  peaks  ($r$,$\phi$)  will appear in the Radon spectra.  Each propagating  crests  in  the  spatiotemporal $\eta(z,t)$  field  is  detected  from  their signatures in the Radon space corresponding to a peak value, where the $\phi$ angle indicates the direction of propagation with respect to the $z$ spatial direction considered. The phase speed of a wave propagating it the $z$ direction will then be obtained using equation~\ref{eq:appendix_RadonPhaseSpeed}. In the case of a spatiotemporal wave field containing incoming ($\eta_{up}$) and outgoing ($\eta_{down}$) waves, such that $\eta(z,t)=\eta_{up}(z,t) +\eta_{down}(z,t)$, each component can be separated using the inverse RT. The inverse RT is a back-projection of $R(r,\phi)$ at given angles $\phi$. The total initial wave signal $\eta(z,t)$ can be reconstructed from the Radon space to the physical space as \citep{almar2014use}

\begin{equation}\label{eq:appendix_inverseRadon}
    \eta = \oiint R(r,\phi) d\phi dr,
\end{equation}

\noindent therefore, the separated wave components can be obtained by applying the limits of integration to the inverse Radon transform as 

\begin{equation}
\begin{aligned}
    & \eta_{up}(z,t) = \int\limits_{-\infty}^{+\infty} \int_{1}^{89} R(r,\phi) d\phi dr, \\
    & \eta_{up}(z,t) = \int\limits_{-\infty}^{+\infty} \int_{91}^{179} R(r,\phi) d\phi dr.
\end{aligned}
\end{equation}

Note that the Radon transforms have been successfully applied for separating wave components in different fields, from as surface or internal ocean wave dynamics, to pressure fluctuation concerning aeroacoustic applications \citep{copeland1995localized,challenor2001use,yoo2011depth,zhang2009wave,martarelli2013subsonic,almar2014use}.

\section{Geometry variations} \label{appendix:LargerCavity}

The RT was also used to separate the upward and downward traveling spirals in the more complicated patterns observed in a cavity four times larger than the previous setup considered (which was based on the experimental cavity presented in~\citet{seelig2018experimental,Meletti2020GAFD}), i.e., instead of an $H=\SI{700}{\milli\metre}$ height cavity, we consider a $H=\SI{2800}{\milli\metre}$ tall cavity. 
\begin{figure}
        \begin{minipage}[t]{0.4\linewidth}
			\centering
			\includegraphics[width=1\linewidth]{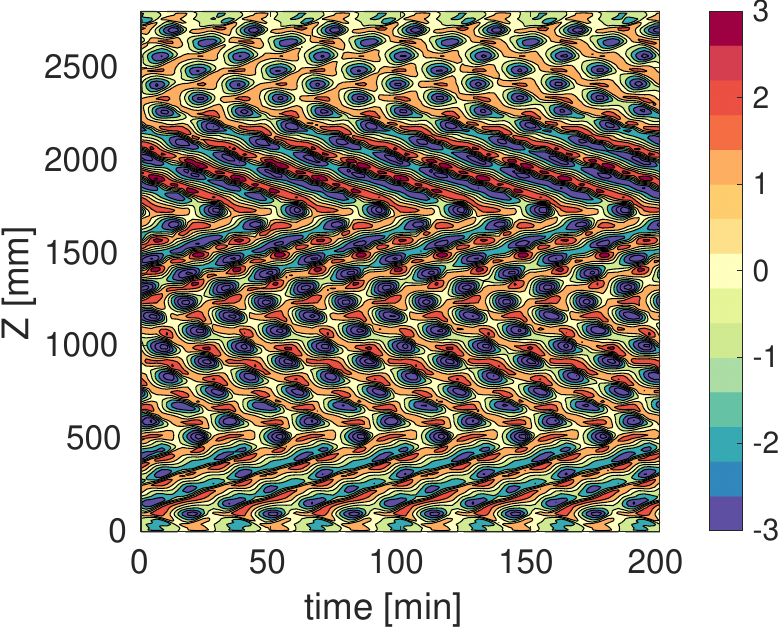}
			\caption*{(a)~Space-time diagram, $H=2.8m$} 
			\vspace{2ex}
	    \end{minipage}
        \begin{minipage}[t]{0.4\linewidth}
			\centering
			\includegraphics[width=1\linewidth]{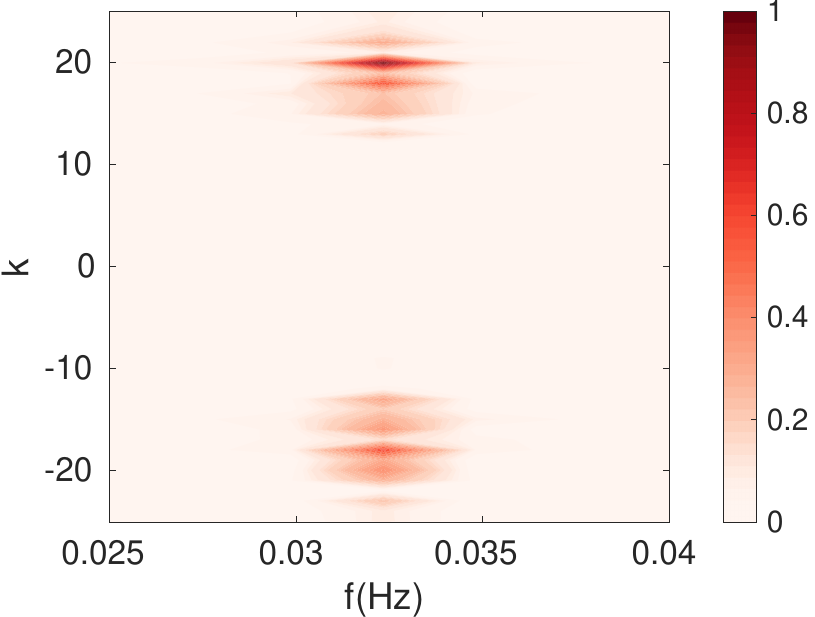}
			\caption*{(b)~2D FFT} 
			\vspace{2ex}
		\end{minipage}
		
		\begin{minipage}[t]{0.4\linewidth}
			\centering
				\includegraphics[width=1\linewidth]{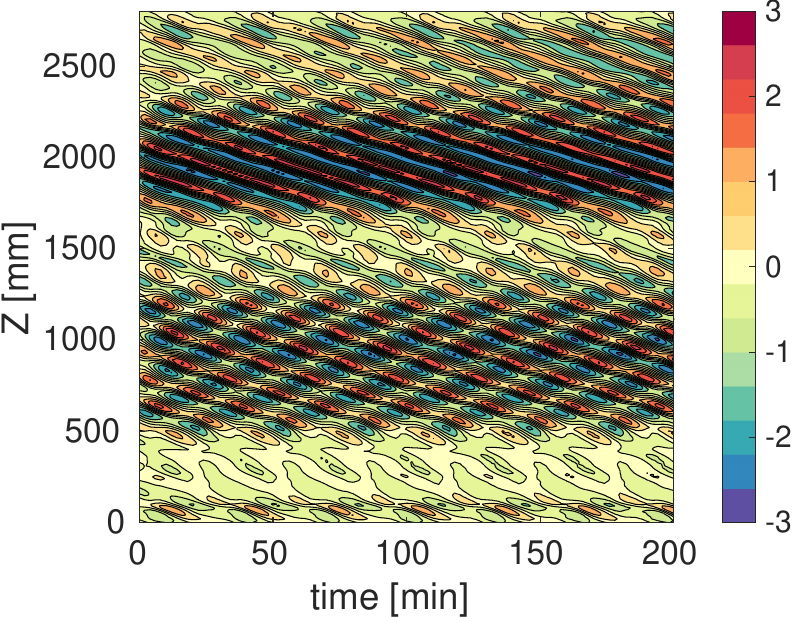}
			\caption*{(c)~Upward component} 
			\vspace{2ex}
		\end{minipage}
		\begin{minipage}[t]{0.4\linewidth}
			\centering							\includegraphics[width=1\linewidth]{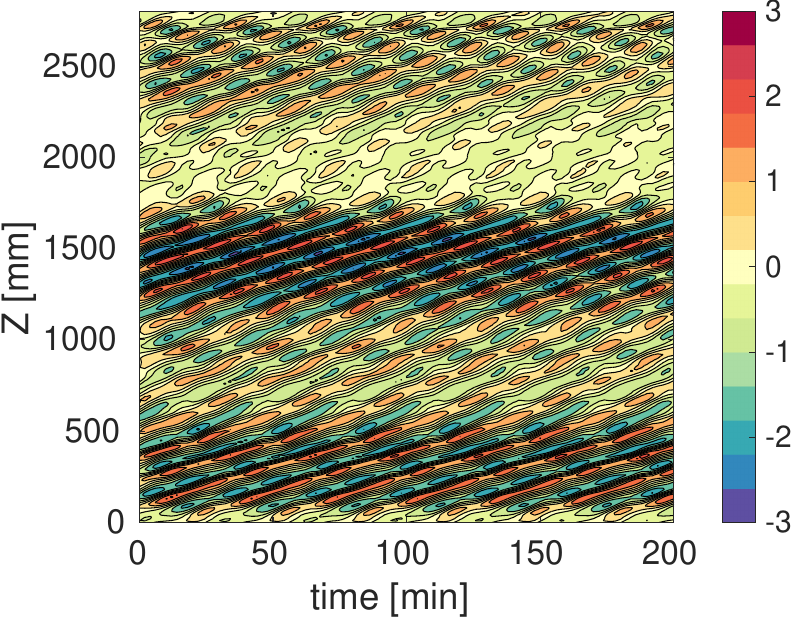}
			\caption*{(d)~Downward component} 
			\vspace{2ex}
		\end{minipage}
		\caption{Separation of upward and downward axial traveling components space-time diagram using the Radon Transform. Results are of $u_\phi$ numerical simulations with $Re=400$, $\mu=0.35$, $\Delta T/\Delta z \approx \SI{5.71}{\kelvin\per\metre}$ and cavity height of $H=\SI{2800}{\milli\metre}$ (four times larger than the previous one considered). (a) Space-time diagram showing the full spiral propagation; (b)~2D-FFT of the full spiral. (c)~Space-time diagram of the spiral component traveling upward; (d)~Space-time diagram of the spiral component traveling downward. \label{fig:Radom_H2p8} }
\end{figure}
On figures~\ref{fig:Radom_H2p8}, the separation of the more complicated patterns in the $H=\SI{2800}{\milli \metre}$ cavity also leads to two spirals with similar wavelengths traveling in opposite axial directions, with the final spiral pattern formed by a linear superposition of these 2 separated components. We highlight that the separation of the spiral components using $u_\phi$ and $u_\phi^\prime = u_\phi-\overline{u_\phi}$ are equivalent since the base flow propagates in the azimuth direction, therefore, it is filtered out by the RT in the axial direction. From figures~\ref{fig:Radom_H2p8}, it becomes clear that the changes in the final spiral direction are associated with the amplitude of each separated axial spiral component since the spiral amplitudes are enhanced in different regions of the axial axis. Note that, when the amplitudes in the upward traveling component are enhanced, the amplitude in the downward component becomes smaller, and the opposite is also true, maintaining constant the energy contained in both amplitudes, with $A_1+A_2=\text{constant}$. Adding the upward (figure~\ref{fig:Radom_H2p8}(c)) and downward (figure~\ref{fig:Radom_H2p8}(d)) spiral components, the initial spiral pattern in figure~\ref{fig:Radom_H2p8}(a) is reconstructed, showing that the spiral patterns arise from the linear superposition of these two upward and downward spiral components with different wave numbers, traveling in time with the same frequencies $\omega$ (in \si{\hertz} in the x-axis of fig.~\ref{fig:Radom_H2p8}.(b)).

The impact of the outer cylinder wall on SRI development was also investigated. In our numerical simulations, the inner cylinder radius $ r_{in} = 75 $ mm was kept constant, while the outer radius $ r_{out} $ was increased ($ r_{out} > 145 $ mm), with all other parameters remaining unchanged. Suppression of the SRI was observed when the outer cylinder radius reached $ r_{out} = 180 $ mm, with only small oscillations occurring at the very beginning of the simulations, which soon vanished into a stable flow with no further development of SRI oscillations after more than 3 hours (in physical time) of simulation.

Although increasing the external radius to $ r_{out} \geq 180 $ mm led to stable SRI flows, when the stratification was increased from $ \Delta T/ \Delta z = \SI{5.71}{\kelvin\per\metre} $ to $ \Delta T/ \Delta z = \SI{11.43}{\kelvin\per\metre} $, SRI oscillations were again observed. These results differ from those observed by \citet{Rudiger2009}, who related to the linear analysis of the SRI, where a wider gap required rather weak stratification to support the SRI, but, in our case, we are changing the aspect ratio of the cavity when we increase the gap size while keeping the height constant. However, they agree with their results considering variations in the Froude number leading to a stable SRI solution. Moreover, we note that changes in $ r_{out} $ led to a delay in the instability development. It is also important to note that the evaluation presented here accounts for the influence of nonlinearities in the simulations, which differs from \citet{Rudiger2009}. Simulations with a slightly increased gap size, from $ d = 70 $mm to $ d = 95 $mm, developed the SRI only after $ t > 50$ minutes. When $ d = 105$~mm and $ \Delta T/ \Delta z = \SI{11.43}{\kelvin\per\metre} $, the time necessary for the first SRI oscillations increased to $ t > 100 $ minutes. Therefore, it is not possible to conclusively say from these investigations whether the instability is suppressed in larger gap sizes (or without an external wall), or if it will simply develop at a later time. When the outer cylinder wall was increased to $ r_{out} = 290 $ mm while maintaining the higher $ \Delta T/ \Delta z = \SI{11.43}{\kelvin\per\metre} $, the SRI oscillations were once again suppressed. Thus, while it is not possible to conclusively determine from this simple qualitative investigation whether the SRI will no longer occur in larger gap widths or if its development is merely delayed, it is clear that the presence of the outer wall can influence the timing of SRI development. A more comprehensive study of the SRI parameters would be important to gain a better understanding of the influence of the outer cylinder wall and critical layers on the development of these instabilities, but our data suggested that critical layers play a significant role in SRI circulation dynamics, which should be further explored in future studies.